%% file: main.tex
\documentclass[adraft]{eptcs}

\usepackage{breakurl}             
\usepackage{underscore}           

\newcommand{\figref}[1]{Figure~\ref{#1}}
\newcommand{\secref}[1]{Section~\ref{#1}}
\newcommand{\tabref}[1]{Table~\ref{#1}}

\newcommand{\algref}[1]{Algorithm~\ref{#1}}

\usepackage[dvipsnames]{xcolor}
\hypersetup{
	hidelinks
}

\usepackage{pgf}
\usepackage{tikz}
\usepackage{amsmath}
\usepackage[ruled,vlined,linesnumbered]{algorithm2e}

\usepackage{amsfonts}

\usetikzlibrary{patterns, shapes.geometric}

\usetikzlibrary{arrows,automata}
\usetikzlibrary{decorations.pathreplacing}
\usetikzlibrary{shapes}
\usetikzlibrary{fit}
\input{railwaytikz}

\usepackage{amsthm}

\theoremstyle{definition}
\newtheorem{definition}{Definition}
\theoremstyle{definition}
\newtheorem{proposition}{Proposition}
\theoremstyle{definition}
\newtheorem{example}{Example}

\colorlet{colort1}{Violet}
\colorlet{colort2}{Peach}
\colorlet{colort3}{RubineRed}

\title{Improving Online Railway Deadlock Detection \\ using a Partial Order Reduction}
\author{Bjørnar Luteberget
\institute{SINTEF Digital AS, Oslo, Norway}
\email{bjornar.luteberget@sintef.no}
}
\def\titlerunning{Improving Online Railway Deadlock Detection using a Partial Order Reduction}
\def\authorrunning{B. Luteberget}
\begin{document}
\maketitle

\begin{abstract}

Although railway dispatching on large national networks is gradually becoming
more computerized, there are still major obstacles to retrofitting
(semi-)autonomous control systems. In addition to requiring extensive and
detailed digitalization of infrastructure models and information systems, exact
optimization for railway dispatching is computationally hard. Heuristic
algorithms and manual overrides are likely to be required for semi-autonomous
railway operations for the foreseeable future.

In this context, being able to detect problems such as deadlocks can be a
valuable part of a runtime verification system. If bound-for-deadlock situations
are correctly recognized as early as possible, human operators will have more
time to better plan for recovery operations. Deadlock detection may also be
useful for verification in a feedback loop with a heuristic or semi-autonomous
dispatching algorithm if the dispatching algorithm cannot itself guarantee a
deadlock-free plan.

We describe a SAT-based planning algorithm for online detection of
bound-for-deadlock situations. The algorithm exploits parallel updates of train
positions and a partial order reduction technique to significantly reduce the
number of state transitions (and correspondingly, the sizes of the formulas) in
the SAT instances needed to prove whether a deadlock situation is bound to
happen in the future. Implementation source code and benchmark instances are
supplied, and a direct comparison against another recent study demonstrates
significant performance gains.

\end{abstract}

\section{Introduction}\label{sec:deadlockrail:introduction}

The discrete elements used in the control of railway traffic, along with the
tightly controlled, predictable environment, makes railway operations a good
match for discrete mathematical modeling formalisms, such as classical planning,
temporal planning, propositional logic, mixed-integer programming, etc.
There is a large amount of literature on railway operations such as routing and
scheduling in both the field of mathematical optimization (see
\cite{handbookrailwayoptimization}, demonstrating the diversity and
sophistication of the field) and in the field of formal methods (see
\cite{fmrailsurvey2} and \cite{fmrailsurvey1}), 
analyzing both for safety and for performance.

Systems for automating re-routing and re-scheduling can have significant impact
on a railway system's overall delay recovery (and consequently also significant
economical impact). However, they have also proven to become intractable at the
network scales required for modeling complex railway systems. Routing and
scheduling algorithms will have the side-effect of detecting deadlocks (because
a bound-for-deadlock system will not have any possible valid schedules), though
that is seldom their main purpose. On-line deadlock detection as a separate
analysis may be useful in the following scenarios:

\begin{itemize}
\item
  In \textbf{manual train dispatching}, trains may enter a
  bound-to-deadlocked state after certain crucial decisions made by a dispatcher. 
  Although this happens very infrequently on
  European railways dominated by relatively short passenger trains, it
  has been known to happen. It is a more pressing issue in 
  the North American railway systems (see \cite{pachldeadlock}), which are
  more dominated by freight trains that may be much
  longer that the usual European case. If deadlocks arise in the actual real-time position of the trains, it
  will be valuable for human operators to be notified of this as early as
  possible, because extraordinary operational modes will typically be
  required to resolve the deadlock by pushing or pulling trains backwards
  out of the deadlocked area.

\item
  In \textbf{semi-automatic or automatic dispatching}, and
  \textbf{semi-autonomous or autonomous train driving},
  it is still unclear whether
  autonomous systems will in practice
  solve (NP-hard) optimization problems exactly, or whether they will
  use heuristic approximations. If the latter is the case, then they
  cannot in general promise deadlock freedom (which is in itself also
  an NP-hard problem, see \cite{DBLP:journals/tomacs/LuDL04}), and detecting future
  deadlock states may be valuable,
  either for use in development, testing, and as a run-time verification tool,
  or for internal verification inside a heuristic optimization algorithm.
  I.e., if deadlocks arise after applying the schedules produced by an
approximate scheduling algorithm, then this result can be fed back into
the scheduling algorithm to help it suggest a new deadlock-free schedule. 
A similar idea was used in \cite{mazzanti,mazzanti2},
where a heuristic algorithm is repaired using model checking to fix all possible deadlock situations 
possibly created for a specific infrastructure.
An approach based on Petri net models solves a similar problem \cite{FANTI20061231},
though neither of these have on-line performance constraints.
Also, in \cite{DBLP:conf/dasc/HasebeTK17}, model checking is used to avoid deadlocks in 
autonomous last-mile transportation.
\end{itemize}

For both of these scenarios, the real-time situation in the physical railway
system changes every few seconds or minutes, so for an analysis algorithm to be
useful it should have an upper limit on typical running times somewhere between
10 seconds and 1 minute. The recent study in \cite{sassoticks} highlights that
there are few effective approaches for determining online (i.e., on real-time
data) whether a running railway system is currently headed towards a deadlocked
state. Although all online planning approaches for solving re-routing and
re-scheduling trains (see \cite{fang} for a survey of these) implicitly solve
the deadlock problem, these approaches usually result in large and sophisticated
optimization problem formulations that can be hard to solve exactly. Whenever
such re-routing and re-scheduling systems resort to heuristic algorithms (such
as in \cite{fengli}) they cannot actually give a definite answer to whether the
system is bound for deadlock. The algorithm presented in \cite{sassoticks} uses
a classical planning model solved with a mixed-integer programming (MIP) solver,
and simplifies other operational aspects  to focus in
on the deadlock detection problem. Their approach calculates a bound on the
number of transitions required and, when instantiating the
model with this number of steps, the infeasibility of the resulting MIP problem
indicates that the system is in a bound-for-deadlock state. However, the
approach is not fully satisfactory because the number of planning steps required
in the worst case makes the problem intractable in practice.

Classical planning problems can be solved by exact mathematical solvers (such as
MIP and SAT solvers) by encoding the state of the system and the transitions
between consecutive states. SAT-based planning is closely related to bounded model checking
(BMC),
and indeed there is a large overlap in the
techniques used for planning as satisfiability and model checking
(see \cite{bmc}, \cite{vizelsatmodelchecking} and
\cite[Ch. 18 and 19]{sathandbook}). 
The number of planning steps used when solving classical planning problems
greatly impacts the performance of the solver (see \cite{rintanen}), but
there is no general way to know how many transitions will be required to solve a
classical planning problem. In some special cases, one can compute reasonable
problem-specific upper (and lower) bounds (called the \emph{completeness
threshold} in BMC) and use that to instantiate the formula used in the
mathematical solver. Another approach is to build the formula incrementally,
adding more steps as required until the desired solution appears (see
\cite{sheeraninduction}, \cite{temporalinduction}).

\input{introexampleimproved}

Parallel plans, i.e.\ allowing multiple actions to take place in the same
transition, is an important technique for reducing the number of steps required
to find a plan (see \cite{rintanen}). However, when the railway system is bound
for a deadlock, having reduced the minimum length of the plan does not
necessarily help much, as it is the \emph{maximum} plan length that determines
how many transitions are required to show a deadlock. In this paper, we show how
to reduce both the minimum and the maximum plan length in a SAT-based planning
model of railway operations by enforcing a \emph{maximal progress constraint}
that not only allows short plans but also forces them to have minimum length by
forcing equivalent partial orders to have the same representation (called
process semantics in \cite{rintanen}, \cite{DBLP:conf/concur/Heljanko01}).

\figref{fig:deadlockrail:introexampleimproved} shows a  
railway with four two-track stations connected by single-track lines.
In the situation shown in State 0, a parallel plan could potentially have been found after only 
4 transitions if the two long trains were short enough to meet at one of the stations, 
although the plan would require up to 26 actions to be executed. 
To show that the system is deadlocked using classical planning without 
the maximal progress constraint would then also require 26 transitions.
Adding the maximal progress constraint that forces all allocations
to happen as early as possible makes the SAT formula 
unsatisfiable (thus proving a bound-for-deadlock situation)
already after 4 transitions.

This paper contains the following contributions:

\begin{itemize} 
	\item a state transition system  encoding into propositional logic, based on our previous work in \cite{lutebergetsatmodsim}
	and adapted for 
	online railway deadlock detection (\secref{sec:deadlockrail:encoding}),
	allowing parallel application of planning operators both for different trains and for contiguous
	sections of a single train's path.
	\item successive algorithmic improvements on the base propositional encoding: 
returning early when finding the shortest valid plan (\secref{sec:deadlockrail:algorithm1}),
applying a global progress constraint and returning early  also when there are no possible actions (\secref{sec:deadlockrail:algorithm2}), and 
applying a maximal progress
constraint that enforces a partial order reduction, further reducing transitions 
(\secref{sec:deadlockrail:algorithm3}).
 \item a demonstration of the
improved performance of our SAT-based solver 
compared against running time and transitions count values from the
literature (specifically \cite{sassoticks}) (\secref{sec:deadlockrail:perf}).

The source code and problem 
instances are available online.

\end{itemize}

\section{Problem Background}\label{sec:deadlockrail:background}

In this paper, we aim
at solving
the online railway deadlock detection problem, i.e.
    given a set of trains and their current locations, we want to determine whether the trains can
	    reach their destinations or if the system is bound for a deadlock state.

The solution to this problem must show that exactly one of these two
conditions hold:

\begin{itemize}
    \item
	      A plan for getting the trains to their destination exists, so there is
		no deadlock.
	\item
		  There cannot exist \emph{any} plan to get the trains to their
		    destinations, so the trains will, by proceeding forward by any paths,
				    eventually end up in a deadlocked state.
\end{itemize}

Note that, in contrast with deadlocks in computer programs, the
goal of railway deadlock analysis is not to prove that there cannot be
any deadlocks.
For a computer program it is usually
interesting to determine whether the program is completely safe from
\emph{any} deadlock occurring.
In railway dispatching, on the other hand, there would usually be many opportunities for a human
dispatching operator to cause a
deadlock. The goal of railway deadlock analysis is to warn, as
early as possible, when the operational situation changes to a state where
there is no way to avoid reaching a deadlock in the future, i.e. the system
is \emph{bound for a deadlocked state}.

In the rest of this section, we describe the relevant domain background for
railway operations and control systems needed to solve the deadlock problem,
and then give a formal problem definition, and finally we give a brief
introduction to SAT-based  planning techniques which 
are used in \secref{sec:deadlockrail:algs}
for solving the problem.

\subsection{Railway Control Systems}

\input{routefig}

A comprehensive model for railway operations analysis takes into account all of the following:

\begin{itemize}
\item
	Track and signalling component layout: railway tracks, switches (branching tracks), 
  	signal locations, detector locations, gradients, curves, tunnels,  etc.
\item
  The interlocking control system: takes commands to send a train from 
  one location to another, and is responsible for allowing only  movements which 
  are safe from collision and derailment.
\item
  Communication constraints: visual signals or radio communication
  are used to tell train drivers that the train may proceed.
\item
  Train characteristics: the trains' lengths, acceleration and braking
  power, maximum speeds, etc. determine how they travel.
\end{itemize}

Fortunately, a much simpler model suffices to prove the absence or presence of a
deadlock state. Since the interlocking only allows trains to move after
exclusively allocating a discrete portion of the infrastructure, using the
interlocking's discretization of the infrastructure into \emph{elementary
routes} (see \figref{fig:scheduling:elementaryroute}), is sufficient to solve
the online deadlock detection problem.
Note that elementary routes must be allocated as a whole, but can be
partially released
  (see
\figref{fig:scheduling:plannerinputstructs}). The parts are called \emph{partial routes}.
The trains' lengths are also relevant for deadlocks, since they
determine whether an allocated route can be freed for use by other
trains after the train has passed on to the next route.

Neither the communication constraints nor a train's dynamic characteristics are
relevant for detecting a deadlock, since the existence of a valid plan does
not depend on the time it takes for the trains to travel (so acceleration,
braking, velocities, and signal sighting distance, are all irrelevant).

We disregard the use of safety zones (known as \emph{overlaps} in British
English, \emph{durchrutschweg} in German) in this paper. Safety zones on
real-world railways are unlikely to cause deadlocks. In many control systems
they are simply freed after a timeout, and durations are irrelevant for the
deadlock problem. Also, safety zone information does not figure in the benchmark
instances we have used (from \cite{sassoticks}). Handling of safety zones in a
SAT-based planning model is described in \cite{lutebergetsatmodsim}.

\input{plannerinputstructs}
\subsection{Problem Definition}

An instance of the online railway deadlock detection problem uses data about the infrastructure, and about the trains and their current locations.
We define the route-based infrastructure model (summarized below as $I$)  as the following data, similar to that used in 
in \cite[Sec. 3.2]{lutebergetsatmodsim}:
\begin{itemize}
	\item A set of partial routes, $\text{PRoutes}$, and a set of elementary routes, $\text{ElemRoutes} \subset 2^{\text{PRoutes}}$, where each partial route belongs to exactly one elementary route.
	\item A set of route delimiters, $\text{Delims}$, each representing either a signal or a detector.
	\item Each partial route's length, $\text{routeLength} : \text{PRoutes} \rightarrow \mathbb{R}$.
	\item Each partial route's entry and exit delimiters, which take the null value at the model boundaries:
		\[ \text{entry},\text{exit} : \text{PRoutes} \rightarrow \text{Delims} \cup \left\{ \text{null} \right\} \]

	\item Conflicts between partial routes, $\text{Conflicts} \subset \text{PRoutes} \times \text{PRoutes}$.
\end{itemize}
In addition, we need the following train data (summarized below as $T$):
\begin{itemize}
	\item A set of trains, $\text{Trains}$.
	\item The length of each train, $\text{trainLength} : \text{Trains} \rightarrow \mathbb{R}$.
	\item Each train's initial position, $\text{initialRoutes} : \text{Trains} \rightarrow 2^\text{PRoutes}$.
	\item Each train's final position alternatives, $\text{finalRoutes} : \text{Trains} \rightarrow 2^{\text{PRoutes}}$.
\end{itemize}
Note that a train that is required to leave the infrastructure at a specific model boundary specifies the 
route leading to that boundary as its final location, i.e. a route $r$ s.t. $\text{exit}(r) = \text{null}$, 
and the train can then leave the model (i.e., it does not need to stay in the route).

A plan $p \in \text{Plans}$ for moving the trains from their initial positions to their final positions is a sequence of pairs of trains
and elementary routes,
\[ \text{Plans} = \left( \text{Trains} \times \text{ElemRoutes} \right)^*, \]
where the $*$ symbol denotes a sequence of elements.
A plan is \emph{correct} if 
it extends each train's initial routes with elementary routes which are
correctly linked by consecutive delimiters ($\text{entry}, \text{exit}$)  to
lead them to one of their final
position alternatives, and
the plan can be executed without becoming \emph{blocked}. A plan execution is blocked if, 
after allowing all trains to reach the end of their currently allocated paths, 
the route in the next step of the plan cannot be allocated because there is a conflict.
Constraints ensuring correctness are described in \secref{sec:deadlockrail:encoding} below.

\begin{definition}
	The online railway deadlock detection problem $D=(I,T)$ is solved by a
	deadlock detection algorithm $d : D \rightarrow \left\{ \text{\textsc{Live}}, \text{\textsc{Dead}} \right\}$, which
	returns $\text{\textsc{Live}}$ if  a \emph{correct plan} exists, and 
	 $\text{\textsc{Dead}}$ otherwise.
\end{definition}
\subsection{SAT-based Planning and Parallel Plans}

A classical planning problem is defined by a set of state
variables \(v \in V\), a set of actions \(a \in A\), an initial state
\(\text{init} : V \rightarrow \mathbb{B}\) and a goal state defined on
some variables \(V_{\text{goal}} \subset V\),
\(\text{goal} : G \rightarrow \mathbb{B}\). For simplicity, we let all variable domains be Boolean ($v \in \mathbb{B}$).
Actions have pre-conditions and post-conditions describing, respectively,
the requirements for an action to take place and the effects of an
action as a set of values assigned to state variables
\(V_{\text{pre}}^a, V_{\text{post}}^a \subset V\),
\(\text{pre} : V_{\text{pre}}^a \rightarrow \mathbb{B}\) and
\(\text{post} : V_{\text{post}}^a \rightarrow \mathbb{B}\).
A solution to the planning problem is a sequence of actions
\(\pi = \left< a_1, a_2, \ldots, a_n \right>\) which transforms the
initial state into the goal state.

A straight-forward encoding of classical planning into SAT
creates a copy of the state space for each planning step, and allows application of 
exactly one action between each pair of consecutive states. In a single
transition, variables that are not modified by the action can not
change, ensuring that the action's pre- and post-conditions are
correctly satisfied. However, this encoding produces a large number of transitions 
when solving problem sizes encountered in real-life applications. If we instead allow multiple
actions to be applied in each step, then we can reduce the number of
steps required and thereby improve solver performance.

The actions in our railway representation are $A = \left\{ a_{t,r} \mid t \in \text{Trains},\, r \in \text{PRoutes} \right\}$, 
meaning that train $t$ allocates route $r$, where the 
preconditions are train consistency constraints and route exclusions,
and post-conditions are occupations \(o_{r}=t\). These conditions are
formalized in the \secref{sec:deadlockrail:encoding} below.

In fact, the total order of actions over-specifies a plan 
because many train movements happen concurrently and only a few of them have 
resource conflicts that makes their ordering meaningfully different.
Starting from a plan $\pi = \left< a_{t_x, r_y}, \ldots \right>$, we define a strict partial order 
$\prec_{\pi}$ as the smallest partial order containing:
\begin{enumerate}
\item each train's path, i.e. if train $t$ takes routes $r_1,r_2,r_3,\ldots...$, 
	then $a_{t,r_1} \prec_\pi  a_{t,r_2},\ \ a_{t,r_2} \prec_\pi a_{t,r_3}$, $\ldots$.
\item for each pair of actions where $a_{t_1,r_1}$ precedes $a_{t_2,r_2}$ in the plan, 
	if $(r_1  ,r_2) \in \text{Conflicts}$, then $a_{t_1,r_1} \prec_\pi a_{t_2,r_2}$.
\end{enumerate}
Two different plans $\pi_1 \neq \pi_2$ producing the same partial order
$\prec_{\pi_1} = \prec_{\pi_2}$ are equivalent for the purposes of deadlock
detection, a fact that we will exploit to reduce state space of the planning
problem.

Following \cite{rintanen}, we consider three approaches to find parallelizable actions,
each corresponding (coincidentally) to one type of parallelization that we have used 
in the railway SAT encoding in \secref{sec:deadlockrail:encoding}:
\begin{itemize}
	\item \textbf{\(\forall\)-steps}:
	When a set of actions can be applied in any order and still produce the same result,
	we may allow this set of actions to take place in the same transition.

	The most easily discovered parallel
  action opportunity we find in the route-based railway model is that
  actions that apply to different trains can always be applied in
  parallel. The mutual exclusion conditions on the route occupation
  variables are enough to make actions for different trains completely
  independently applicable, so this is a \(\forall\)-steps set of actions. 

		\item \textbf{\(\exists\)-steps}:
When a set of actions applied between two states is guaranteed to have at least one total order that 
transforms the first state into the second state,
	we may allow this set of actions to take place in the same transition.
		This is also called the \emph{post-serializability} condition,
		and is potentially a very general framework for parallelizing
		planning actions.
		
		In the route-based railway model we exploit the following
		opportunity for \(\exists\)-steps parallelism: 
		considering each train in the separately, if we
  have a train $t$ that travels on the sequence of routes \(r_1, r_2, r_3\),
		the corresponding actions $a_{t,r_1}, a_{t,r_2}, a_{t,r_3}$ must apply in the same sequence. However,
  there is no problem in applying all these actions in the same step
  without representing the choice of order. Given a set of connected
  routes in an acyclic infrastructure, there is exactly one possible
  ordering of the actions, namely the one given by the directed route
  graph. So a set of routes forming a train path can be applied in
  parallel and \emph{post-serialized} into a set of actions.
  Trains spanning long paths is also used in \cite{DBLP:journals/scp/HongHP17} for model checking, 
  though there to abstract away train lengths.

		\begin{figure}[b]
  \begin{center} \begin{tikzpicture}[->,>=stealth',shorten >=1pt,auto,xscale=1.75,yscale=0.5]

	\node[anchor=south] at (0,0.5) {Step 1};
	\node[anchor=south] at (1,0.5) {Step 2};
	\node[anchor=south] at (2,0.5) {Step 3};

	\node (a1) at (0,0) {$\alpha_1$};
	\node (a2) at (0,-1) {$\alpha_2$};
	\node (a3) at (1,-1) {$\alpha_3$};
	\node (a4) at (2,-1) {$\alpha_4$};
	\path 
	(a1) edge (a4)
	(a2) edge (a3)
	(a3) edge (a4);

	\draw[-] (3,1) -- (3,-1.5);

	\begin{scope}[shift={(4,0)}]
		
	\node[anchor=south] at (0,0.5) {Step 1};
	\node[anchor=south] at (1,0.5) {Step 2};
	\node[anchor=south] at (2,0.5) {Step 3};

	\node (a1) at (1,0) {$\alpha_1$};
	\node (a2) at (0,-1) {$\alpha_2$};
	\node (a3) at (1,-1) {$\alpha_3$};
	\node (a4) at (2,-1) {$\alpha_4$};
	\path 
	(a1) edge (a4)
	(a2) edge (a3)
	(a3) edge (a4);
	\end{scope}
\end{tikzpicture} \end{center}
\caption{Two different state transition system plans for the actions $\alpha_1,\alpha_2,\alpha_3,\alpha_4$. The arrows show the
 dependencies between actions. Whether $\alpha_1$ is executed in step 1 or 2 results in the same partial order.}
\label{fig:deadlockrail:processsemanticsreduction}
\end{figure}

		\item \textbf{Process semantics reduction}:
		Consider a classical planning
  problem with four actions $\alpha_1$, $\alpha_2$, $\alpha_3$, $\alpha_4$, where \(\alpha_2,\alpha_3,\alpha_4\) must be
  executed in order, while \(\alpha_1\) only needs to precede \(\alpha_4\).
  Solving this problem with SAT-based planning requires only three steps (not four), since we
  have parallel action application of \(\alpha_1\) with either \(\alpha_2\) or \(\alpha_3\).
  However, \(\alpha_1\) can be still be applied either in the first step or the
  second step, and both describe the same partial order of actions (see \figref{fig:deadlockrail:processsemanticsreduction}).

We can remove these different but equivalent representations from the SAT problem by simply
constraining actions to happen as early as possible. 
The encoding is described in \secref{sec:deadlockrail:algorithm3} below.
This property is called process semantics in \cite{rintanen}, 
\cite{DBLP:conf/concur/Heljanko01}, and can have a 
significant impact on planning solver performance.
Adding such a constraint, which we here call the \emph{maximal progress constraint},
removes redundant plan representations that reduce to the same partial order $\prec$.

  This constraint does not add
  parallelism between actions to reduce the \emph{minimum} number of
  steps required to reach a state, but instead reduces the \emph{maximum} number of steps
  required to execute a plan.
  For unsatisfiable planning problems (such as in a bound-for-deadlock state) 
  it is a particular advantage to avoid a high maximum number of steps with 
  little progress in each step.

  We used this maximal progress constraint in previous work (\cite{lutebergetsatmodsim}),
  where we enumerated all plans, although there we still used a heuristic for choosing 
  the upper bound on the number of transitions.

\end{itemize}

\section{SAT Encoding and Algorithms}
\label{sec:deadlockrail:algs}

In this section, we first present a SAT encoding 
of route-based railway as a state transition system (\secref{sec:deadlockrail:encoding}).
Then we develop an algorithm for the online deadlock
detection problem, using this SAT encoding, presented as three successively improving algorithms.
First, a simple SAT-based planning algorithm with a statically 
computed upper bound on steps (\secref{sec:deadlockrail:algorithm1}),
then we add a modification to detect deadlocks before reaching 
the static upper bound on steps (\secref{sec:deadlockrail:algorithm2}),
and finally we add the maximal progress constraint that further reduces the 
number of steps until deadlock (\secref{sec:deadlockrail:algorithm3}).

\subsection{State transition encoding}
\label{sec:deadlockrail:encoding}
We now encode a state transition representation of the  deadlock detection problem
as propositional logic
 (a simplified and adapted version of the encoding in \cite{lutebergetsatmodsim}, ignoring safety zones).
We write the propositional logic formula for a $k$-step unrolling of the transition relation as
\( \Phi_k = \bigwedge_{i=0}^{k} \phi_i\),
where $\phi_0$ is the known initial state.
We use well-known SAT encoding techinques for at-most-one constraints and
one-hot encoded finite sets  
		(see e.g. \cite{DBLP:conf/cp/Sinz05}, \cite[Ch. 2]{sathandbook} or \cite{bjorksatencoding} 
  for common encoding techniques and their tradeoffs).
The formula $\phi_i$ for each state $i \geq 1$ of the system uses the following state variables:

\begin{enumerate}
\item Each partial route $r$ in step $i$ has an \textbf{occupancy status} $o^i_{r}$ which
  is either free ($o^i_{r} = \text{Free}$) or occupied by a specific
  train $t$ ($o^i_{r} = t$). We use a one-hot encoding to represent the finite set as Booleans
  so each $o_r^i=t$ is a 
  variable in the SAT instance.
\item Each train $t$ in step $i$ has a Boolean representing \textbf{finished
    status} $f^i_t$, signifying that the train reaches its destination in 
		  state $i$ or before.
\end{enumerate}

Note here that we do not need to add additional variables
for each action $a_{t,r}$ in step $i$ (which we write $a_{t,r}^i$)
when encoding this problem to conjunctive normal form:
for any clause \(C\), we can make the clause conditional on \(a_{t,r}^i\) by writing
\(\left( (o^{i-1}_{r}) \neq t \wedge (o^i_{r} =t)  \right) \Rightarrow C \)  as the equivalent \( (o^{i-1}_{r} = t) \vee 
(o^i_{r} \neq t) \vee C\),
which is a clause.

The following constraints apply to states and pairs of consecutive states to encode a \emph{correct} plan:
\begin{itemize}
\item \textbf{Each train takes a single, contiguous path.}

First, ensure that only one route from a given start delimiter may be taken at
any time:

\[
C_1^i = \bigwedge_{t \in \text{Trains}} 
\bigwedge_{d \in \text{Delims} 
		}	
\text{atMostOne}( \left\{ o_r^i=t \mid \text{entry(r)} = d \right\})
\]

Routes can only be allocated to a train when they
extend some other route which was already allocated to the same train, i.e.,
consecutive routes must match so that the exit delimiter of one is the entry delimiter of the next:

\[
	C_2^i = 
	\bigwedge_{t \in \text{Trains}}
	\bigwedge_{\substack{a \in \text{PRoutes}}}
	(
		a_{t,a}^i
		\Rightarrow
		\bigvee_{\substack{b \in \text{PRoutes} \\\text{entry}(a) = \text{exit}(b) }} o^{i}_{b} = t
	)
	\]

Note that this constraint ensures that the trains' allocation to routes
\emph{locally}
forms a path in the graph of routes. 
In an acyclic infrastructure, this is sufficient. An approach for handling cyclic infrastructure is described in \cite{lutebergetsatmodsim}.

\item \textbf{Conflicting routes are not active simultaneously.}

\[ C_3^i = \bigwedge_{(a,b) \in \text{Conflicts}} \left( (o_a^i=\text{Free}) \vee (o_b^i=\text{Free}) \right) \]

\item \textbf{Elementary routes are allocated as a unit.}

The partial release feature of the interlocking system is handled by 
splitting each elementary route into
separate routes for each component which can be released separately.
The set $\text{ElemRoutes}$ contains such sets of routes.
Partial routes forming an elementary route must be allocated together:

\[	C_4^i = \bigwedge_{t \in \text{Trains}}	
\bigwedge_{e \in \text{ElemRoutes}}
\bigwedge_{r \in e}
\left(
a_{t,r}^i \Rightarrow \bigwedge_{r \in e } (o_r^i=t)
\right)
\]

\item \textbf{Partial routes are only deactivated after a train has fully passed over them.}

Routes are freed when sufficient length has been allocated ahead to fully contain the train.

\[ C_5^i = \bigwedge_{t \in \text{Trains}} \bigwedge_{r \in \text{PRoutes}}
 (o_r^i = t) \Rightarrow \left((o_r^{i+1} \neq t ) \Leftrightarrow \text{freeable}_t^i(r, \text{trainLength}(t))\right)  \]

Note that the bidirectional implication sign on the right hand side means that deallocation is both allowed and required.
The leftward part of this implication simplifies 
the train path consistency constraints  because we do not need constraints to prevent trains from deallocating a section in the 
middle of their path and allocating an alternative path instead.
The freeable formulas are produced by the following recursive function:
\[
\begin{aligned}
	\text{freeable}_t^i(a,l) &= \text{\textbf{if }}  \text{exit}(a) = \text{null} \text{\textbf{ or }} l \le 0 \text{\textbf{ then }} \top \text{\textbf{ else }} \\ 
    \\
      & \bigvee_{\substack{b \in \text{PRoutes} \\ \text{exit}(a) = \text{entry}(b)}}
        (o_b^i=t) \wedge 
            \text{freeable}_t^i(b,l - \text{routeLength}(a))
	. \\
        \end{aligned}
    \]

 The function returns a disjunction of possible routes that can be taken after
 route $a$, i.e.\ from delimiter $\text{exit}(a)$,
 but if those routes are shorter than the length of the train, then they must be conjuncted with
 the freeable formula for the following route $b$.
 Note that the $\text{freeable}$ function itself is not part of the SAT problem, but
 is used to compute a formula based on the static infrastructure data. 
 No numerical representations are used in the SAT formulas.

\end{itemize}

Additionally, setting the finished status requires visiting one of the final routes (and $f_t^0= \bot $):
\[
	C_6^i = \bigwedge_{t \in \text{Trains}} 
	\left(
	(\neg f_t^{i-1} \wedge f_t^i ) \Rightarrow 
	\bigvee_{r \in \text{finalRoutes}(t)} o_r^i=t
	\right)
	\]

Combining these constraints, we get the state transition representation of state $i$ as 
\[\phi_i = C_1^i \wedge C_2^i \wedge C_3^i \wedge  C_4^i \wedge C_5^i \wedge C_6^i.\]
Finally, reaching the goal state at state $i$ (or before) is encoded as
\( G_i =
\bigwedge_{t \in \text{Trains}} f_t^i\).

\subsection{Algorithm 1: Upper-bounding \texorpdfstring{$k$}{k}}
\label{sec:deadlockrail:algorithm1}

Completeness thresholds (upper-bounding the number of transitions) are often too 
large for practical use, but for the online railway deadlock detection we require a 
complete algorithm, so we compute a completeness threshold here as a starting point for 
further algorithms.  Let $n_j^k$ be the number of elementary routes on the longest path
(recall that the infrastructure is assumed to be acyclic) from the starting
position of train $t_j$ to its $k$th alternative destination in a specific
problem instance $D=(I,T)$.  Let $U = \sum_j \max_k n_j^k$. 
\begin{proposition}
If any correct plan for $D$ exists, $\Phi_U \wedge G_U$ must be satisfiable. 
\end{proposition}

This is indeed the approach taken in \cite{sassoticks}, where a lower and an
upper bound on the number of planning steps is computed, by computing the lower
and upper bound on the number of steps a single train in isolation would need
to proceed to one of its destinations.  
However, always instantiating the system with $U$ planning
steps is computationally expensive, and indeed \cite{sassoticks} 
explores a varying number of transitions to see if a plan can be found with fewer transitions.
\algref{alg:deadlockrail:a1} implements a similar strategy where all $i=1,2,\ldots$ are
tried successively, adding $\phi_i$'s to the SAT solver incrementally.
The $G_i$'s are added temporarily using the \emph{assumptions} interface that many 
incremental SAT solvers support.

\begin{algorithm}
\DontPrintSemicolon
\SetAlgoLined
\SetKwInOut{Input}{Input}\SetKwInOut{Output}{Output}
	\Input{A problem instance $D=(I,T)$ and a bound $k$.}
	\Output{$\text{\textsc{Dead}}$ if the system is bound for deadlock, $\text{\textsc{Live}}$ otherwise. }
	\textbf{let} $i=1$. \\
	\textbf{if} $\Phi_i \wedge G_i$ is \textsc{Sat}, \textbf{return} \textsc{Live} \\
	\textbf{if} $i < k$, increment $i$ and go to $2$, \textbf{else} \textbf{return} \textsc{Dead} \\
\caption{Online railway deadlock detection using incremental $k$-bounded model checking}\label{alg:deadlockrail:a1}
\end{algorithm}

\subsection{Algorithm 2: Early Return on Deadlocks}
\label{sec:deadlockrail:algorithm2}
Instead of detecting deadlocks implicitly by the 
unsatisfiability of the formula
when the number of transitions has reached the upper bound, 
we can decrease the number of required transitions by adding an explicit
global progress constraint that requires that each step of the plan makes
some amount of progress. In any planning step, even if it is the first one, if the
progress constraint cannot be satisfied, then the system must be bound for a
deadlock state.

At least one route must be allocated in each transition:

\[
z_i = 
\bigvee_{t \in \text{Trains}}
\bigvee_{r \in \text{PRoutes}} a_{t,r}^i,\quad \quad \quad 
	Z_i = \bigwedge_{j=1}^i z_j
\]

\begin{proposition}
If $\Phi_i \wedge Z_i$ is unsatisfiable for any $i$, then the instance is \textsc{Dead}.
\end{proposition}

\algref{alg:deadlockrail:a2}
improves on 
\algref{alg:deadlockrail:a1}
by not needing to solve
all $\Phi_i$ for $i$ all the way up to the explicitly computed completeness
threshold $U$, but can instead terminate at the lowest number of steps where we
can be sure that the system is in fact bound to be deadlocked. This is similar
to the inductive proof algorithms from \cite{sheeraninduction}, but using incremental 
SAT like the  ``zig-zag'' algorithm from \cite{temporalinduction}.

\begin{algorithm}[h]
\DontPrintSemicolon
\SetAlgoLined
\SetKwInOut{Input}{Input}\SetKwInOut{Output}{Output}
	\Input{A problem instance $D=(I,T)$.}
	\Output{$\text{\textsc{Dead}}$ if the system is bound for deadlock, $\text{\textsc{Live}}$ otherwise. }
	\textbf{let} $i=1$. \\
	\textbf{if} $\Phi_i \wedge Z_i$ is \textsc{Unsat}, \textbf{return} \textsc{Dead} \\
	\textbf{if} $\Phi_i \wedge Z_i \wedge G_i$ is \textsc{Sat}, \textbf{return} \textsc{Live} \\
	increment $i$ and go to $2$. \\
\caption{Online railway deadlock detection with global progress constraint}\label{alg:deadlockrail:a2}
\end{algorithm}

\begin{example}
	Consider the following situation, where both trains are trying to exit
	at the model boundary in their traveling direction.

	
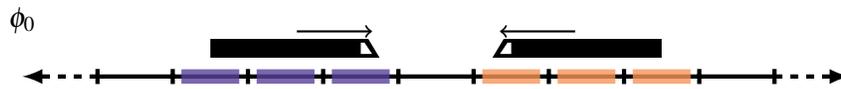
\begin{center}
		\vspace{0.5em}
	\begin{tikzpicture}
		\node at (-1,0.75) {$\phi_0$};
		\draw[rail] 
			(0,0) \railNSend -- 
			(1,0) \railNSend -- 
			(2,0) \railNSend -- 
			(3,0) \railNSend -- 
			(4,0) \railNSend -- 
			(5,0) \railNSend -- 
			(6,0) \railNSend -- 
			(7,0) \railNSend -- 
			(8,0) \railNSend -- 
			(9,0) \railNSend;
		\draw[rail, ->, >=latex, dashed] (0,0) -- (-1.0,0);
		\draw[rail, ->, >=latex, dashed] (9,0) -- (10.0,0);
		\trainoverarrow{1.5}{0}{2.25}{1.0}
		\trainoverarrowleft{7.5}{0}{2.25}{1.0}
		\draw[shorten >=0.3em, shorten <=0.3em,line width=0.5em,color=colort1,opacity=0.8] (1,0) -- (2,0);
		\draw[shorten >=0.3em, shorten <=0.3em,line width=0.5em,color=colort1,opacity=0.8] (2,0) -- (3,0);
		\draw[shorten >=0.3em, shorten <=0.3em,line width=0.5em,color=colort1,opacity=0.8] (3,0) -- (4,0);
		\draw[shorten >=0.3em, shorten <=0.3em,line width=0.5em,color=colort2,opacity=0.8] (5,0) -- (6,0);
		\draw[shorten >=0.3em, shorten <=0.3em,line width=0.5em,color=colort2,opacity=0.8] (6,0) -- (7,0);
		\draw[shorten >=0.3em, shorten <=0.3em,line width=0.5em,color=colort2,opacity=0.8] (7,0) -- (8,0);
	\end{tikzpicture}
		\vspace{0.5em}
	\end{center}

The situation is obviously bound for deadlock. 
However, because each train can take up to 5 steps to reach its model boundary,
	Algorithm 1 would require solving the transition system using $U = 10$ steps (corresponding
	to the sum of the length of the paths for each train) before concluding that the situation is \textsc{Dead}.
	Algorithm 2 would return \textsc{Dead} already after adding 2 transitions.
	The first transition would require either train to move ahead into the unoccupied
	route between them, and the second transition would then have no possible 
	actions, so 
	the formula $\Phi_2 \wedge Z_2$ is unsatisfiable.

	
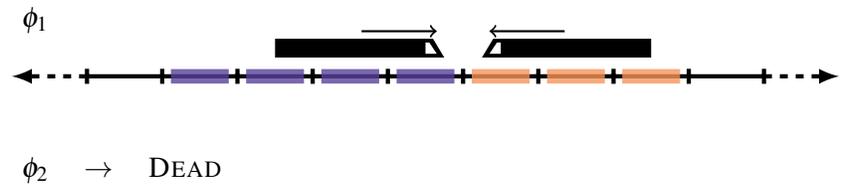
\begin{center}
		\vspace{0.5em}
	\begin{tikzpicture}
		\node [anchor=west]at (-1,0.75) {$\phi_1$};
		\node [anchor=west]at (-1,0.75-2) {$\phi_2 \quad \rightarrow \quad \text{\textsc{Dead}}$};
		\draw[rail] 
			(0,0) \railNSend -- 
			(1,0) \railNSend -- 
			(2,0) \railNSend -- 
			(3,0) \railNSend -- 
			(4,0) \railNSend -- 
			(5,0) \railNSend -- 
			(6,0) \railNSend -- 
			(7,0) \railNSend -- 
			(8,0) \railNSend -- 
			(9,0) \railNSend;
		\draw[rail, ->, >=latex, dashed] (0,0) -- (-1.0,0);
		\draw[rail, ->, >=latex, dashed] (9,0) -- (10.0,0);
		\trainoverarrow{2.5}{0}{2.25}{1.0}
		\trainoverarrowleft{7.5}{0}{2.25}{1.0}
		\draw[shorten >=0.3em, shorten <=0.3em,line width=0.5em,color=colort1,opacity=0.8] (1,0) -- (2,0);
		\draw[shorten >=0.3em, shorten <=0.3em,line width=0.5em,color=colort1,opacity=0.8] (2,0) -- (3,0);
		\draw[shorten >=0.3em, shorten <=0.3em,line width=0.5em,color=colort1,opacity=0.8] (3,0) -- (4,0);
		\draw[shorten >=0.3em, shorten <=0.3em,line width=0.5em,color=colort1,opacity=0.8] (4,0) -- (5,0);
		\draw[shorten >=0.3em, shorten <=0.3em,line width=0.5em,color=colort2,opacity=0.8] (5,0) -- (6,0);
		\draw[shorten >=0.3em, shorten <=0.3em,line width=0.5em,color=colort2,opacity=0.8] (6,0) -- (7,0);
		\draw[shorten >=0.3em, shorten <=0.3em,line width=0.5em,color=colort2,opacity=0.8] (7,0) -- (8,0);
	\end{tikzpicture}
		\vspace{0.5em}
	\end{center}

\end{example}

This is only the best case, though, there are still simple cases where the
number of steps required to prove a deadlock still scales proportionally
to the $k$ value in Algorithm 1.

\begin{example}
	Consider the following situation (similar to Example 1):

	
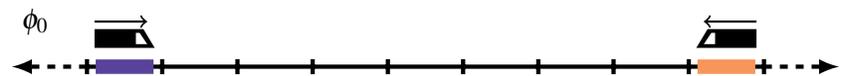
\begin{center} \vspace{0.5em} \begin{tikzpicture}
		\begin{scope}[shift={(0,0)}]
		\node [anchor=west]at (-1,0.6) {$\phi_0$};
		\draw[rail] (0,0) \railNSend -- (1,0) \railNSend -- (2,0) \railNSend -- (3,0) \railNSend -- (4,0) \railNSend -- (5,0) \railNSend -- (6,0) \railNSend -- (7,0) \railNSend -- (8,0) \railNSend -- (9,0) \railNSend;
		\draw[rail, ->, >=latex, dashed] (0,0) -- (-1.0,0); \draw[rail, ->, >=latex, dashed] (9,0) -- (10.0,0);
		\trainoverarrow{0.1}{0}{0.8}{0.7} \trainoverarrowleft{8.9}{0}{0.8}{0.7}
		\draw[shorten >=0.3em, shorten <=0.3em,line width=0.5em,color=colort1] (0,0) -- (1,0);
		\draw[shorten >=0.3em, shorten <=0.3em,line width=0.5em,color=colort2] (8,0) -- (9,0);
		\end{scope}
	\end{tikzpicture} \vspace{0.5em} \end{center}

	This situation is obviously deadlocked in the same way as Example 1. However, here, Algorithm 2 
	requires 7 transition steps to terminate.

	
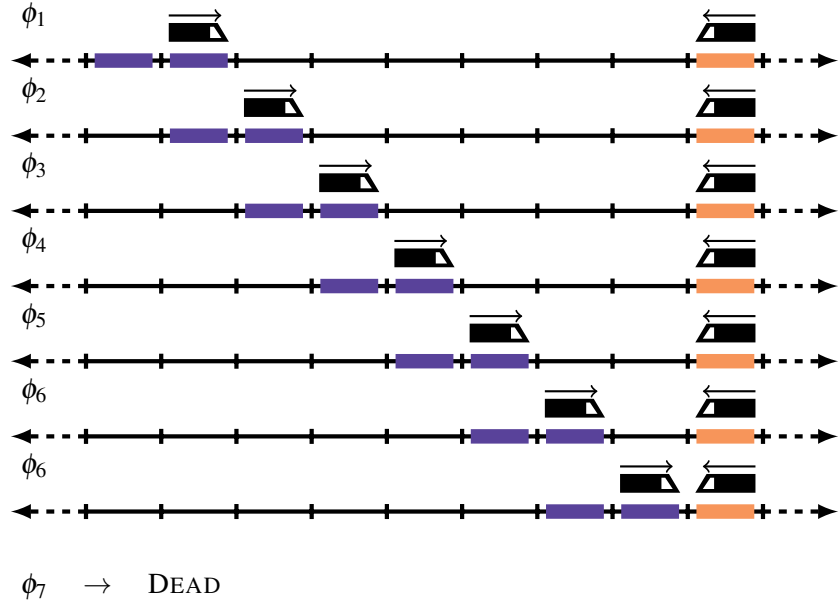
\begin{center} \vspace{0.5em} \begin{tikzpicture}
		\begin{scope}[shift={(0,-1.0)}]
		\node [anchor=west]at (-1,0.6) {$\phi_1$};
		\draw[rail] (0,0) \railNSend -- (1,0) \railNSend -- (2,0) \railNSend -- (3,0) \railNSend -- (4,0) \railNSend -- (5,0) \railNSend -- (6,0) \railNSend -- (7,0) \railNSend -- (8,0) \railNSend -- (9,0) \railNSend;
		\draw[rail, ->, >=latex, dashed] (0,0) -- (-1.0,0); \draw[rail, ->, >=latex, dashed] (9,0) -- (10.0,0);
		\trainoverarrow{1.1}{0}{0.8}{0.7} \trainoverarrowleft{8.9}{0}{0.8}{0.7}
		\draw[shorten >=0.3em, shorten <=0.3em,line width=0.5em,color=colort1] (0,0) -- (1,0);
		\draw[shorten >=0.3em, shorten <=0.3em,line width=0.5em,color=colort1] (1,0) -- (2,0);
		\draw[shorten >=0.3em, shorten <=0.3em,line width=0.5em,color=colort2] (8,0) -- (9,0);
		\end{scope}
		\begin{scope}[shift={(0,-2.0)}]
		\node [anchor=west]at (-1,0.6) {$\phi_2$};
		\draw[rail] (0,0) \railNSend -- (1,0) \railNSend -- (2,0) \railNSend -- (3,0) \railNSend -- (4,0) \railNSend -- (5,0) \railNSend -- (6,0) \railNSend -- (7,0) \railNSend -- (8,0) \railNSend -- (9,0) \railNSend;
		\draw[rail, ->, >=latex, dashed] (0,0) -- (-1.0,0); \draw[rail, ->, >=latex, dashed] (9,0) -- (10.0,0);
		\trainoverarrow{2.1}{0}{0.8}{0.7} \trainoverarrowleft{8.9}{0}{0.8}{0.7}
		\draw[shorten >=0.3em, shorten <=0.3em,line width=0.5em,color=colort1] (1,0) -- (2,0);
		\draw[shorten >=0.3em, shorten <=0.3em,line width=0.5em,color=colort1] (2,0) -- (3,0);
		\draw[shorten >=0.3em, shorten <=0.3em,line width=0.5em,color=colort2] (8,0) -- (9,0);
		\end{scope}
		\begin{scope}[shift={(0,-3.0)}]
		\node [anchor=west]at (-1,0.6) {$\phi_3$};
		\draw[rail] (0,0) \railNSend -- (1,0) \railNSend -- (2,0) \railNSend -- (3,0) \railNSend -- (4,0) \railNSend -- (5,0) \railNSend -- (6,0) \railNSend -- (7,0) \railNSend -- (8,0) \railNSend -- (9,0) \railNSend;
		\draw[rail, ->, >=latex, dashed] (0,0) -- (-1.0,0); \draw[rail, ->, >=latex, dashed] (9,0) -- (10.0,0);
		\trainoverarrow{3.1}{0}{0.8}{0.7} \trainoverarrowleft{8.9}{0}{0.8}{0.7}
		\draw[shorten >=0.3em, shorten <=0.3em,line width=0.5em,color=colort1] (2,0) -- (3,0);
		\draw[shorten >=0.3em, shorten <=0.3em,line width=0.5em,color=colort1] (3,0) -- (4,0);
		\draw[shorten >=0.3em, shorten <=0.3em,line width=0.5em,color=colort2] (8,0) -- (9,0);
		\end{scope}
		\begin{scope}[shift={(0,-4.0)}]
		\node [anchor=west]at (-1,0.6) {$\phi_4$};
		\draw[rail] (0,0) \railNSend -- (1,0) \railNSend -- (2,0) \railNSend -- (3,0) \railNSend -- (4,0) \railNSend -- (5,0) \railNSend -- (6,0) \railNSend -- (7,0) \railNSend -- (8,0) \railNSend -- (9,0) \railNSend;
		\draw[rail, ->, >=latex, dashed] (0,0) -- (-1.0,0); \draw[rail, ->, >=latex, dashed] (9,0) -- (10.0,0);
		\trainoverarrow{4.1}{0}{0.8}{0.7} \trainoverarrowleft{8.9}{0}{0.8}{0.7}
		\draw[shorten >=0.3em, shorten <=0.3em,line width=0.5em,color=colort1] (3,0) -- (4,0);
		\draw[shorten >=0.3em, shorten <=0.3em,line width=0.5em,color=colort1] (4,0) -- (5,0);
		\draw[shorten >=0.3em, shorten <=0.3em,line width=0.5em,color=colort2] (8,0) -- (9,0);
		\end{scope}
		\begin{scope}[shift={(0,-5.0)}]
		\node [anchor=west]at (-1,0.6) {$\phi_5$};
		\draw[rail] (0,0) \railNSend -- (1,0) \railNSend -- (2,0) \railNSend -- (3,0) \railNSend -- (4,0) \railNSend -- (5,0) \railNSend -- (6,0) \railNSend -- (7,0) \railNSend -- (8,0) \railNSend -- (9,0) \railNSend;
		\draw[rail, ->, >=latex, dashed] (0,0) -- (-1.0,0); \draw[rail, ->, >=latex, dashed] (9,0) -- (10.0,0);
		\trainoverarrow{5.1}{0}{0.8}{0.7} \trainoverarrowleft{8.9}{0}{0.8}{0.7}
		\draw[shorten >=0.3em, shorten <=0.3em,line width=0.5em,color=colort1] (4,0) -- (5,0);
		\draw[shorten >=0.3em, shorten <=0.3em,line width=0.5em,color=colort1] (5,0) -- (6,0);
		\draw[shorten >=0.3em, shorten <=0.3em,line width=0.5em,color=colort2] (8,0) -- (9,0);
		\end{scope}
		\begin{scope}[shift={(0,-6.0)}]
		\node [anchor=west]at (-1,0.6) {$\phi_6$};
		\draw[rail] (0,0) \railNSend -- (1,0) \railNSend -- (2,0) \railNSend -- (3,0) \railNSend -- (4,0) \railNSend -- (5,0) \railNSend -- (6,0) \railNSend -- (7,0) \railNSend -- (8,0) \railNSend -- (9,0) \railNSend;
		\draw[rail, ->, >=latex, dashed] (0,0) -- (-1.0,0); \draw[rail, ->, >=latex, dashed] (9,0) -- (10.0,0);
		\trainoverarrow{6.1}{0}{0.8}{0.7} \trainoverarrowleft{8.9}{0}{0.8}{0.7}
		\draw[shorten >=0.3em, shorten <=0.3em,line width=0.5em,color=colort1] (5,0) -- (6,0);
		\draw[shorten >=0.3em, shorten <=0.3em,line width=0.5em,color=colort1] (6,0) -- (7,0);
		\draw[shorten >=0.3em, shorten <=0.3em,line width=0.5em,color=colort2] (8,0) -- (9,0);
		\end{scope}
		\begin{scope}[shift={(0,-7.0)}]
		\node [anchor=west]at (-1,0.6) {$\phi_6$};
		\draw[rail] (0,0) \railNSend -- (1,0) \railNSend -- (2,0) \railNSend -- (3,0) \railNSend -- (4,0) \railNSend -- (5,0) \railNSend -- (6,0) \railNSend -- (7,0) \railNSend -- (8,0) \railNSend -- (9,0) \railNSend;
		\draw[rail, ->, >=latex, dashed] (0,0) -- (-1.0,0); \draw[rail, ->, >=latex, dashed] (9,0) -- (10.0,0);
		\trainoverarrow{7.1}{0}{0.8}{0.7} \trainoverarrowleft{8.9}{0}{0.8}{0.7}
		\draw[shorten >=0.3em, shorten <=0.3em,line width=0.5em,color=colort1] (6,0) -- (7,0);
		\draw[shorten >=0.3em, shorten <=0.3em,line width=0.5em,color=colort1] (7,0) -- (8,0);
		\draw[shorten >=0.3em, shorten <=0.3em,line width=0.5em,color=colort2] (8,0) -- (9,0);
		\end{scope}
		\begin{scope}[shift={(0,-8.0)}]
		\node [anchor=west]at (-1,0.0) {$\phi_7 \quad \rightarrow \quad \text{\textsc{Dead}}$};
		\end{scope}
	\end{tikzpicture} \vspace{0.5em} \end{center}

\end{example}

Note here that if the two trains were on separate tracks 
(so that they were not bound for deadlock), then the instance would
have been proved \textsc{Live} using only one transition.
The parallel actions that allow arriving at the \textsc{Live} result
in fewer transitions does not correspondingly allow arriving at the \textsc{Dead} result
in similarly few transitions, as this example shows, because the \emph{minimum} amount of progress in each
transition is small.

\subsection{Algorithm 3: Collapsing Equivalent Partial Orders}
\label{sec:deadlockrail:algorithm3}
Looking in more detail at the actions taken by each train at each time step, we
see that the global progress constraint $Z_i$ can be satisfied
by trains moving forward by only one route per transition, as Example 2
demonstrates. 
We can 
force more progress to happen in each step
using the following insight:
for any $a_{t,r}$ taking place in state $i$ (except for $i=1$),
a route conflicting with $r$ must have been occupied in the previous state.
If there was no such conflict in the previous state, performing $a_{t,r}$
in state $i-1$ would be possible, and would  produce the same partial order $\prec$.

More precisely (and taking into account that the distinction between elementary routes and partial routes),
we define the maximal progress constraint $P_i = \bigwedge_{j=2}^i p_j$, where:

\[
	p_i = 
	\bigwedge_{t \in \text{Trains}}
	\bigwedge_{\substack { e \in \text{ElemRoutes} \\ r \in e }}
\left(
	\left( a_{t,r}^i \wedge 
	\bigvee_{\substack{x \in \text{PRoutes}  \\ \text{entry}(r) = \text{exit}(x) }} (o_x^{i-1}=t)
	\right) 
	\Rightarrow
	\bigvee_{\substack{
		y \in e \\ z \in \text{PRoutes} \\ (y,z) \in \text{Conflicts}
		}} \left( o_z^{i-1} \neq t \wedge o_z^{i-1} \neq \text{Free}  \right)
	\right)
	\]

$p_i$ encodes that if route $r$ is allocated to train $t$ in step $i$, and any of the preceding routes $x$ 
were allocated in the previous step, then at least one route conflicting with $r$ must have
been allocated to another train in the previous step. $P_i$ is the conjunction of $p_i$'s from the 
initial state and up to state $i$.
Note that $p_i$ does not apply for $i=1$, because 
the initial state has known constant values for all variables, and the trains
are not necessarily waiting for a route to become unallocated in the initial state.

	\begin{proposition}
The formula $\Phi_i \wedge Z_i \wedge P_i \wedge G_i$ admits all the same plans as $\Phi_i \wedge Z_i \wedge G_i$, modulo the plan partial order, 
i.e.\ adding the maximal progress constraint does not eliminate any valid plans.
	\end{proposition}

	To see why this is the case, consider a plan where, in step $i$, $p_i$ does not hold for a route $r$ allocated to a train $t$.
	This means that no route conflicting with $r$ was allocated (to another train) in state $i-1$.
	To fulfill $p_i$, we can simply allocate $r$ to $t$ in state $i-1$ instead of state $i$, which results in the same
	plan modulo the plan partial order. By repairing all solutions to $\Phi_k$ in this way, we 
	see that adding the constraint $P_k$ preserves all plans modulo the partial order.
	See \cite[Sec. 2.2.]{rintanen} for a more general and thorough treatment of this property.

	\begin{example}
Consider the following scenario:

	
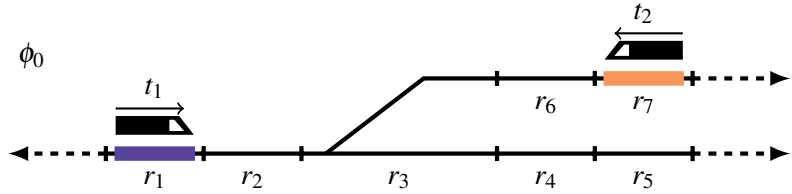
\begin{center} \vspace{0.5em} \begin{tikzpicture}[xscale=1.3]
		\begin{scope}[shift={(0,0)}]
		\node [anchor=west]at (-2,1.3) {$\phi_0$};
			\draw[rail] (-1,0)\railNSend -- (0,0) \railNSend -- (1,0) \railNSend -- (3,0) \railNSend -- (4,0) \railNSend -- (5,0) \railNSend ;
			\draw[rail] (1.25,0) -- (2.25,1) -- (3,1) \railNSend -- (4,1) \railNSend -- (5,1) \railNSend;

		\draw[rail, ->, >=latex, dashed] (-1,0) -- (-2,0); 
		\draw[rail, ->, >=latex, dashed] (5,0) -- (6,0);
		\draw[rail, ->, >=latex, dashed] (5,1) -- (6,1);
		\trainoverarrow{-1+0.1}{0}{0.8}{0.7} \trainoverarrowleft{4.9}{1}{0.8}{0.7}
		\draw[shorten >=0.3em, shorten <=0.3em,line width=0.5em,color=colort1] (0-1,0) -- (1-1,0);
		\draw[shorten >=0.3em, shorten <=0.3em,line width=0.5em,color=colort2] (4,1) -- (5,1);
			\node [anchor=south] at (-0.5,0.6) {$t_1$};
			\node [anchor=south] at (4.5,1.6) {$t_2$};
			\node [anchor=north] at (-0.5,0-0.1) {$r_1$};
			\node [anchor=north] at (0.5, 0-0.1) {$r_2$};
			\node [anchor=north] at (2,   0-0.1) {$r_3$};
			\node [anchor=north] at (3.5, 0-0.1) {$r_4$};
			\node [anchor=north] at (4.5, 0-0.1) {$r_5$};
			\node [anchor=north] at (3.5, 1-0.1) {$r_6$};
			\node [anchor=north] at (4.5, 1-0.1) {$r_7$};
		\end{scope}
	\end{tikzpicture} \vspace{0.5em} \end{center}

	Note that here we use the same route names for both travel directions, for simplicity. 
The following plan would satisfy the formula $\Phi_i \wedge Z_i$ used in
Algorithm 2 with $i = 3$:

	
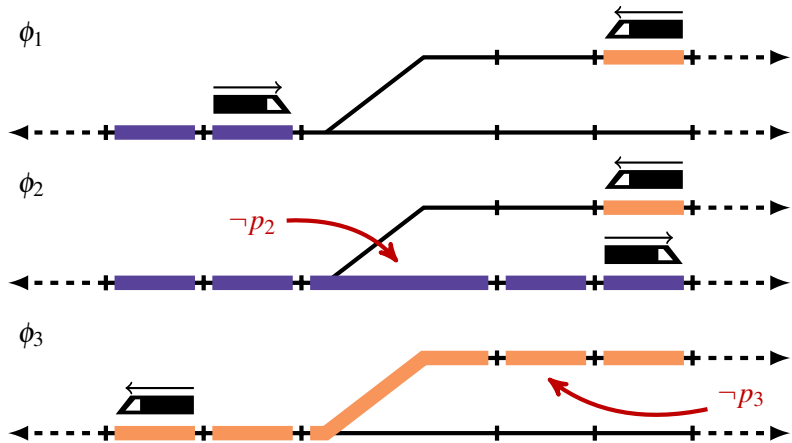
\begin{center} \vspace{0.5em} \begin{tikzpicture}[xscale=1.3]
		\begin{scope}[shift={(0,0)}]
		\node [anchor=west]at (-2,1.3) {$\phi_1$};
			\draw[rail] (-1,0)\railNSend -- (0,0) \railNSend -- (1,0) \railNSend -- (3,0) \railNSend -- (4,0) \railNSend -- (5,0) \railNSend ;
			\draw[rail] (1.25,0) -- (2.25,1) -- (3,1) \railNSend -- (4,1) \railNSend -- (5,1) \railNSend;

		\draw[rail, ->, >=latex, dashed] (-1,0) -- (-2,0); \draw[rail, ->, >=latex, dashed] (5,0) -- (6,0); \draw[rail, ->, >=latex, dashed] (5,1) -- (6,1);
		\trainoverarrow{0.1}{0}{0.8}{0.7} \trainoverarrowleft{4.9}{1}{0.8}{0.7}
		\draw[shorten >=0.3em, shorten <=0.3em,line width=0.5em,color=colort1] (0-1,0) -- (1-1,0);
		\draw[shorten >=0.3em, shorten <=0.3em,line width=0.5em,color=colort2] (4,1) -- (5,1);
		\draw[shorten >=0.3em, shorten <=0.3em,line width=0.5em,color=colort1] (0,0) -- (1,0);
		\end{scope}

		\begin{scope}[shift={(0,-2)}]
		\node [anchor=west]at (-2,1.3) {$\phi_2$};
			\draw[rail] (-1,0)\railNSend -- (0,0) \railNSend -- (1,0) \railNSend -- (3,0) \railNSend -- (4,0) \railNSend -- (5,0) \railNSend ;
			\draw[rail] (1.25,0) -- (2.25,1) -- (3,1) \railNSend -- (4,1) \railNSend -- (5,1) \railNSend;

		\draw[rail, ->, >=latex, dashed] (-1,0) -- (-2,0); \draw[rail, ->, >=latex, dashed] (5,0) -- (6,0); \draw[rail, ->, >=latex, dashed] (5,1) -- (6,1);
		\trainoverarrow{4.1}{0}{0.8}{0.7} \trainoverarrowleft{4.9}{1}{0.8}{0.7}
		\draw[shorten >=0.3em, shorten <=0.3em,line width=0.5em,color=colort1] (0-1,0) -- (1-1,0);
		\draw[shorten >=0.3em, shorten <=0.3em,line width=0.5em,color=colort1] (0,0) -- (1,0);
		\draw[shorten >=0.3em, shorten <=0.3em,line width=0.5em,color=colort1] (1,0) -- (3,0);
		\draw[shorten >=0.3em, shorten <=0.3em,line width=0.5em,color=colort1] (3,0) -- (4,0);
		\draw[shorten >=0.3em, shorten <=0.3em,line width=0.5em,color=colort1] (4,0) -- (5,0);
		\draw[shorten >=0.3em, shorten <=0.3em,line width=0.5em,color=colort2] (4,1) -- (5,1);
			\node[color=red!75!black,anchor=south] (x) at (0.5,0.5) {$\neg p_2$};
			\path[color=red!75!black, ultra thick,->,>=stealth',shorten >=1pt] (x) [bend left] edge (2.0,0.25);
		\end{scope}

		\begin{scope}[shift={(0,-4)}]
		\node [anchor=west]at (-2,1.3) {$\phi_3$};
			\draw[rail] (-1,0)\railNSend -- (0,0) \railNSend -- (1,0) \railNSend -- (3,0) \railNSend -- (4,0) \railNSend -- (5,0) \railNSend ;
			\draw[rail] (1.25,0) -- (2.25,1) -- (3,1) \railNSend -- (4,1) \railNSend -- (5,1) \railNSend;

		\draw[rail, ->, >=latex, dashed] (-1,0) -- (-2,0); \draw[rail, ->, >=latex, dashed] (5,0) -- (6,0); \draw[rail, ->, >=latex, dashed] (5,1) -- (6,1);
			\trainoverarrowleft{0.9-1}{0}{0.8}{0.7}
		\draw[shorten >=0.3em, shorten <=0.3em,line width=0.5em,color=colort2] (3,1) -- (4,1);
		\draw[shorten >=0.3em, shorten <=0.3em,line width=0.5em,color=colort2] (4,1) -- (5,1);
			\draw[shorten >=0.3em, shorten <=0.3em,line width=0.5em,color=colort2] (1,0)--(1.25,0)--(2.25,1)--(3,1);
		\draw[shorten >=0.3em, shorten <=0.3em,line width=0.5em,color=colort2] (0,0) -- (1,0);
		\draw[shorten >=0.3em, shorten <=0.3em,line width=0.5em,color=colort2] (-1,0) -- (-1+1,0);
			\node[color=red!75!black,anchor=south] (x) at (5.5,0.2) {$\neg p_3$};
			\path[color=red!75!black, ultra thick,->,>=stealth',shorten >=1pt] (x) [bend left] edge (3.5,0.75);
		\end{scope}
	\end{tikzpicture} \vspace{0.5em} \end{center}

	There are two violations of $p_i$ here, indicated by the red arrows.
	The first one in state 2, where the rightwards-headed train allocates the branching track when it was already 
	free in $\phi_1$.
	The second one in state 3, where the leftwards-headed train allocates the route between its initial position 
	and the branching route, when it was already free in $\phi_1$.
	An equivalent plan fulfilling $P_k$ uses only two planning steps:

	
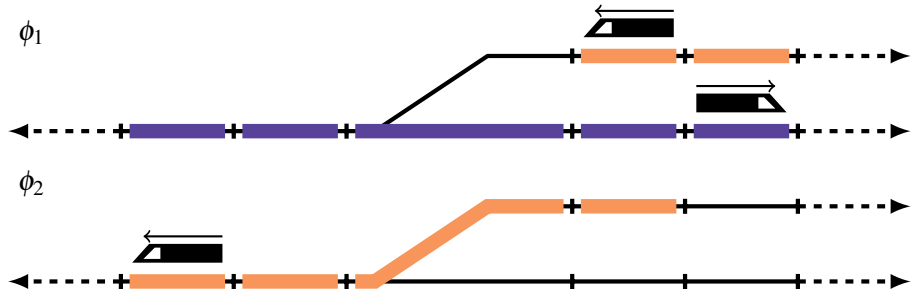
\begin{center} \vspace{0.5em} \begin{tikzpicture}[xscale=1.5]
		\begin{scope}[shift={(0,-2)}]
		\node [anchor=west]at (-2,1.3) {$\phi_1$};
			\draw[rail] (-1,0)\railNSend -- (0,0) \railNSend -- (1,0) \railNSend -- (3,0) \railNSend -- (4,0) \railNSend -- (5,0) \railNSend ;
			\draw[rail] (1.25,0) -- (2.25,1) -- (3,1) \railNSend -- (4,1) \railNSend -- (5,1) \railNSend;

		\draw[rail, ->, >=latex, dashed] (-1,0) -- (-2,0); \draw[rail, ->, >=latex, dashed] (5,0) -- (6,0); \draw[rail, ->, >=latex, dashed] (5,1) -- (6,1);
		\trainoverarrow{4.1}{0}{0.8}{0.7} \trainoverarrowleft{3.9}{1}{0.8}{0.7}
		\draw[shorten >=0.3em, shorten <=0.3em,line width=0.5em,color=colort1] (0-1,0) -- (1-1,0);
		\draw[shorten >=0.3em, shorten <=0.3em,line width=0.5em,color=colort1] (0,0) -- (1,0);
		\draw[shorten >=0.3em, shorten <=0.3em,line width=0.5em,color=colort1] (1,0) -- (3,0);
		\draw[shorten >=0.3em, shorten <=0.3em,line width=0.5em,color=colort1] (3,0) -- (4,0);
		\draw[shorten >=0.3em, shorten <=0.3em,line width=0.5em,color=colort1] (4,0) -- (5,0);
		\draw[shorten >=0.3em, shorten <=0.3em,line width=0.5em,color=colort2] (3,1) -- (4,1);
		\draw[shorten >=0.3em, shorten <=0.3em,line width=0.5em,color=colort2] (4,1) -- (5,1);
		\end{scope}

		\begin{scope}[shift={(0,-4)}]
		\node [anchor=west]at (-2,1.3) {$\phi_2$};
			\draw[rail] (-1,0)\railNSend -- (0,0) \railNSend -- (1,0) \railNSend -- (3,0) \railNSend -- (4,0) \railNSend -- (5,0) \railNSend ;
			\draw[rail] (1.25,0) -- (2.25,1) -- (3,1) \railNSend -- (4,1) \railNSend -- (5,1) \railNSend;

		\draw[rail, ->, >=latex, dashed] (-1,0) -- (-2,0); \draw[rail, ->, >=latex, dashed] (5,0) -- (6,0); \draw[rail, ->, >=latex, dashed] (5,1) -- (6,1);
			\trainoverarrowleft{0.9-1}{0}{0.8}{0.7}
		\draw[shorten >=0.3em, shorten <=0.3em,line width=0.5em,color=colort2] (3,1) -- (4,1);
			\draw[shorten >=0.3em, shorten <=0.3em,line width=0.5em,color=colort2] (1,0)--(1.25,0)--(2.25,1)--(3,1);
		\draw[shorten >=0.3em, shorten <=0.3em,line width=0.5em,color=colort2] (0,0) -- (1,0);
		\draw[shorten >=0.3em, shorten <=0.3em,line width=0.5em,color=colort2] (-1,0) -- (-1+1,0);
		\end{scope}
	\end{tikzpicture} \vspace{0.5em} \end{center}

	Both of these solutions produce the same partial order $\prec$.
	We can visualize the partial order
	by drawing each member  $(\alpha,\beta)$
	of the transitive reduction of the partial order relation
	using lines with arrows pointing from the first component $\alpha$ to the second component $\beta$.
	Then both of the plans in the example above produce the following partial order:

	\begin{center} \vspace{0.5em} \begin{tikzpicture}[->,>=stealth',shorten >=1pt,auto,xscale=1.8]
		\node (n11) at (0,0)  { \Large $a_{t_1, r_1}$ };
		\node (n12) at (1,0)  { \Large $a_{t_1, r_2}$ };
		\node (n13) at (2,0)  { \Large $a_{t_1, r_3}$ };
		\node (n14) at (3,0)  { \Large $a_{t_1, r_4}$ };
		\node (n15) at (4,0)  { \Large $a_{t_1, r_5}$ };
		\node (n27) at (1,-1) { \Large $a_{t_2, r_7}$ };
		\node (n26) at (2,-1) { \Large $a_{t_2, r_6}$ };
		\node (n23) at (3,-1) { \Large $a_{t_2, r_3}$ };
		\node (n22) at (4,-1) { \Large $a_{t_2, r_2}$ };
		\node (n21) at (5,-1) { \Large $a_{t_2, r_1}$ };
		\path 
		(n11) edge (n12)
		(n12) edge (n13)
		(n13) edge (n14)
		(n14) edge (n15)
		(n13) edge (n23)
		(n27) edge (n26)
		(n26) edge (n23)
		(n23) edge (n22)
		(n22) edge (n21)
		;
	\end{tikzpicture} \vspace{0.5em} \end{center}

	\end{example}

In general, adding the constraint $P_k$ forces a particular plan partial order to be
represented in a unique way in the state space.
Such constraints are known
in the BMC literature
as \emph{partial order reductions}.
Importantly, this constraint reduces the \emph{maximum} number of steps that a
set of trains can use to complete the possible set of plans, allowing deadlocks
to become evident using a lower number of planning steps.

	\begin{proposition}
		If $\Phi_i \wedge Z_i \wedge P_i$ is unsatisfiable for any $i$, then the system is \textsc{Dead}.
	\end{proposition}

\algref{alg:deadlockrail:a3} is defined in the same way as \algref{alg:deadlockrail:a2}, only with $\Phi_i$ replaced with $\Phi_i \wedge P_i$.

\begin{algorithm}[h]
\DontPrintSemicolon
\SetAlgoLined
\SetKwInOut{Input}{Input}\SetKwInOut{Output}{Output}
	\Input{A problem instance $D=(I,T)$.}
	\Output{$\text{\textsc{Dead}}$ if the system is bound for deadlock, $\text{\textsc{Live}}$ otherwise. }
	\textbf{let} $i=1$. \\
	\textbf{if} $\Phi_i \wedge Z_i\wedge P_i $ is \textsc{Unsat}, \textbf{return} \textsc{Dead} \\
	\textbf{if} $\Phi_i \wedge Z_i\wedge P_i \wedge G_i$ is \textsc{Sat}, \textbf{return} \textsc{Live} \\
	increment $i$ and go to $2$. \\
\caption{Online railway deadlock detection with partial order reduction}\label{alg:deadlockrail:a3}
\end{algorithm}

\begin{samepage}
	To demonstrate how $P_i$ changes the scaling properties of the deadlock detection algorithm in the best case,
	we have performed a scaling test using two trains starting on opposite sides of an infrastructure
	consisting of $n$ two-track stations, where none of the station routes are long enough for the trains
	to cross:

	\begin{center} \vspace{0.5em} \begin{tikzpicture}

		\begin{scope}[shift={(0,0)},xscale=0.8]
		\node [anchor=west]at (-1,1.3) {$\phi_0$};
			\draw[rail] 
			(0,0)\railNSend -- 
			(1,0)\railNSend -- 
			(2.5,0)\railNSend --
			(3.5,0)\railNSend --
			(5,0)\railNSend --
			(6,0)\railNSend;

			\draw[rail] (1.25,0) -- (2.25,1) -- (2.5,1) \railNSend -- (3.5,1) \railNSend -- (3.75,1) -- (4.75,0);
		\draw[rail, ->, >=latex, dashed] (0,0) -- (-1,0); 
		\draw[shorten >=0.3em, shorten <=0.3em,line width=0.5em,color=colort1] (0,0) -- (1,0);
		\end{scope}
		\begin{scope}[shift={(6*0.8,0)}]
		\draw [decorate,decoration={brace,amplitude=10pt}] (0,0.6) -- node[midway,above,yshift=10pt]{$n$ stations} (3,0.6);
			\draw[rail, dashed, color=gray] (0,0) -- (1,0);
			\node[anchor=center] at (1.5,0.25) {\huge $\ldots$};
			\draw[rail, dashed, color=gray] (2,0) -- (3,0);
		\end{scope}
		\begin{scope}[shift={(6*0.8+3,0)},xscale=0.8]
			\draw[rail] 
			(0,0)\railNSend -- 
			(1,0)\railNSend -- 
			(2.5,0)\railNSend --
			(3.5,0)\railNSend --
			(5,0)\railNSend --
			(6,0)\railNSend;

			\draw[rail] (1.25,0) -- (2.25,1) -- (2.5,1) \railNSend -- (3.5,1) \railNSend -- (3.75,1) -- (4.75,0);
		\draw[rail, ->, >=latex, dashed] (6,0) -- (7,0);
		\draw[shorten >=0.3em, shorten <=0.3em,line width=0.5em,color=colort2] (5,0) -- (6,0);
		\end{scope}
		\trainoverarrow{-1-0.1}{0}{1.8}{1.0} 
		\trainoverarrowleft{3+6*0.8+7.4*0.8}{0}{1.8}{1.0}

	\end{tikzpicture} \vspace{0.5em} \end{center}
\end{samepage}

	Running time results and number of steps for Algorithms 1, 2, and 3, for instances 
	with increasing $n$ are shown in 
	\tabref{tab:deadlockrail:twotrainscaling}.
	We see that, in fact, for any number of stations and possible paths, 3 planning steps
	are sufficient to prove a deadlock involving two trains.

\input{twotraintable}
\input{perftable}

\section{Performance Evaluation} \label{sec:deadlockrail:perf} We have used
the problem instances from \cite{sassoticks}, and
have evaluated algorithms 1, 2, and 3 described in this paper, comparing
them to the reported worst-case number of transitions (called ``\emph{ticks}''
in \cite{sassoticks}) and running time results in \cite{sassoticks}. The results
are shown in \tabref{tab:deadlockrail:perftable}. The running times for
Algorithms 1, 2, 3 were measured on a laptop running Linux on an Intel i7-5500U
CPU using the CaDiCaL SAT solver~\cite{BiereFazekasFleuryHeisinger-SAT-Competition-2020-solvers}. We compare our
algorithms to the \emph{reported} performance from \cite{sassoticks} instead of
executing a complete comparison test on the same machine because the source code
and the closed-source commercial MIP solver library used in \cite{sassoticks}
were not available to us. This means that the comparison is not completely
valid, as differences in hardware, solver libraries, and the heuristic for
number of transitions may influence the result. Note also that the
algorithm in \cite{sassoticks} computes more than just the \textsc{Dead/Live}
status, also suggesting a set of safe locations to park trains for recovery
operations. Still, we claim that the marked difference in steps and running
times between \cite{sassoticks} and our Algorithm 3 shows that our improvements
to a large extent solve the challenges from that paper's conclusion, and opens
the possibility for adequate performance also on larger-scale problem instances.
The implementation source code for algorithms 1, 2, 3, and the problem
instances, are available
online\footnote{\url{https://github.com/luteberget/deadlockrail}}.

\section{Conclusions and Future Work}
We have demonstrated how to use an incremental SAT solver to determine whether
a railway system is bound for a deadlocked state.  By allowing multiple trains
to move by multiple elementary routes for every transition, we greatly lower
the number of transitions required to show that the system is \emph{not}
bound-for-deadlock.  By adding a constraint that forces all resource
allocations to happen immediately after a conflicting resource has been
released, we greatly lower the maximum number of steps where trains can make
meaningful progress, 
allowing us to show that that the system \emph{is} bound-for-deadlock in fewer steps.
Together, these features result in a smaller propositional logic formula to be solved
by the SAT solver because fewer transitions are required to decide the \textsc{Dead/Live} answer, 
resulting in a much more efficient
algorithm for the online railway deadlock detection problem.
Efficient deadlock detection can be useful in non-autonomous decision support systems for 
manual railway dispatching or in 
semi-autonomous or autonomous controllers for railway dispatching or train driving.

We plan to further investigate the possible integration between
non-complete/heuristic analysis algorithms and complete analysis algorithms for
specific critical properties, in the context of autonomous railway operations.
Studying the transition system described in this paper is not only relevant for
deadlock detection in manual or autonomous railway control, but the efficiency
gains described in this paper may also be applied to other analysis tasks in
railway design and operations.  We plan to investigate the
grouping together adjacent of infrastructure elements and/or sets of trains to
hide parts of the problem instances, and then performing
counter-example-guided abstraction refinement. This might further reduce the
sizes of the SAT instances required in complete planning algorithms for railways applications.

\section*{Acknowledgements}

We thank Veronica Dal Sasso for supplying benchmark problem instances and
helping with the interpretation of the data. We thank Carlo Mannino for comments
on the ideas and comparison of our approach with the literature on railway
deadlocks.

\bibliographystyle{eptcs}
\bibliography{bibs}
\end{document}

%% file: railwaytikz.tex
\tikzstyle{sig}=[thick]
\tikzstyle{rail}=[ultra thick]

\newcommand*{\railNSend}{+ (0,.1) -- + (0,-.1) + (0,0)}

\newcommand{\trainoverarrow}[4] {
    \begin{scope}[shift={(#1,#2)}]
        \fill [color=black] (0.0, 0.25) -- (#3,0.25) -- (#3-0.15,0.5) -- (0.0,0.5) -- cycle;
        \fill [color=white] (#3-0.25,0.3) -- (#3-0.1,0.3) -- (#3-0.2,0.45) -- (#3-0.25,0.45) --  cycle;
        \draw[->,thick] (#3-0.10-#4, 0.60) -- (#3-0.10,0.60);
    \end{scope}
}
\newcommand{\trainoverarrowleft}[4] {
    \begin{scope}[shift={(#1,#2)},xscale=-1]
        \fill [color=black] (0.0, 0.25) -- (#3,0.25) -- (#3-0.15,0.5) -- (0.0,0.5) -- cycle;
        \fill [color=white] (#3-0.25,0.3) -- (#3-0.1,0.3) -- (#3-0.2,0.45) -- (#3-0.25,0.45) --  cycle;
        \draw[->,thick] (#3-0.10-#4, 0.60) -- (#3-0.10,0.60);
    \end{scope}
}

\newcommand{\signalsmall}[2] {
        \begin{scope}[shift={(#1,#2)}, scale=0.1]
        \draw [sig] (0,-1) -- (0,1);
        \draw [sig](0,0) -- (2,0);
        \draw [sig](3,0) circle (1);
        \end{scope}
}

                \tikzset{station/.pic = {

                    \draw[fill=black] (1.5,0) -- (1.75,0) -- (1.75,0.25);
                    \draw[fill=black] (5.5,0) -- (5.75,0) -- (5.75,-0.25);
                    \draw[fill=black] (7.5,0) -- (7.25,0) -- (7.25,0.25);

                    \draw[rail] (-0.5,0) -- (0,0) \railNSend -- (1,0) \railNSend -- (1.5,0);
                    \draw[rail] (1.5,0) -- (2.5,1) -- (3.0,1) \railNSend -- (5.0,1) \railNSend -- (6.0,1) \railNSend -- (6.5,1) -- (7.5,0);
                    \draw[rail] (1.5,0) -- (3.0,0) \railNSend -- (5.0,0) \railNSend -- (5.5,0);
                    \draw[rail] (5.5,0) -- (6.5,-1) -- (7,-1) \railNSend -- (9,-1);
                    \draw[draw=none,fill=black] (9.0,-0.8) rectangle (9.1,-1.2);

                    \draw[rail] (5.5,0) -- (6.0,0) \railNSend -- (8,0) \railNSend -- (9,0) \railNSend -- (9.5,0);

                    \draw[->,>=latex,ultra thick] (-0.5,0) --(-0.9,0);
                    \draw[->,>=latex,ultra thick] (9.5,0) --(9.9,0);

                    \signalsmall{0}{-0.3}
                    \signalsmall{5}{-0.3}
                    \signalsmall{5}{1.3}
                    \begin{scope}[rotate=180]
                    \signalsmall{-3}{-1.3}
                    \signalsmall{-3}{0.3}
                    \signalsmall{-9}{-0.3}
                    \signalsmall{-7}{1.0-0.3}
                    \end{scope}
                }}

%% file: introexampleimproved.tex
\begin{figure}[b!]
	\begin{center} 
		\resizebox{\textwidth}{!}
		{
		\begin{tikzpicture}
			\tikzstyle{occ}=[shorten >=0.3em, shorten <=0.3em,line width=0.5em]

\begin{scope}[xscale=0.75,yscale=0.75]
		\node [anchor=west]at (-1,1.3) {State 0};

		\draw[rail] (0,0) \railNSend -- (1,0) \railNSend -- (2.5,0) \railNSend -- (3.5,0) \railNSend -- (5,0) \railNSend ;
		\draw[rail] (1.25,0) -- (2.25,1) -- (2.5,1) \railNSend -- (3.5,1) \railNSend -- (3.75,1) -- (4.75,0);
	\begin{scope}[shift={(5,0)}]
		\draw[rail] (0,0) \railNSend -- (1,0) \railNSend -- (2.5,0) \railNSend -- (3.5,0) \railNSend -- (5,0) \railNSend ;
		\draw[rail] (1.25,0) -- (2.25,1) -- (2.5,1) \railNSend -- (3.5,1) \railNSend -- (3.75,1) -- (4.75,0);
	\end{scope}
	\begin{scope}[shift={(10,0)}]
		\draw[rail] (0,0) \railNSend -- (1,0) \railNSend -- (2.5,0) \railNSend -- (3.5,0) \railNSend -- (5,0) \railNSend ;
		\draw[rail] (1.25,0) -- (2.25,1) -- (2.5,1) \railNSend -- (3.5,1) \railNSend -- (3.75,1) -- (4.75,0);
	\end{scope}
	\begin{scope}[shift={(15,0)}]
		\draw[rail] (0,0) \railNSend -- (1,0) \railNSend -- (2.5,0) \railNSend -- (3.5,0) \railNSend -- (5,0) \railNSend ;
		\draw[rail] (1.25,0) -- (2.25,1) -- (2.5,1) \railNSend -- (3.5,1) \railNSend -- (3.75,1) -- (4.75,0);
	\end{scope}
	\draw[rail, ->, >=latex, dashed] (0,0) -- (-1,0); 
		\draw[rail, ->, >=latex, dashed] (20,0) -- (21,0);

		\trainoverarrow{0.1+0.5}{0}{1.8}{1.0} 
		\trainoverarrowleft{8.4}{1}{0.8}{0.7}
		\trainoverarrowleft{19.4}{0}{1.8}{1.0}
		\draw[occ,color=colort1] (0,0) -- (1,0);
		\draw[occ,color=colort1] (1,0) -- (2.5,0);
		\draw[occ,color=colort2] (17.5,0) -- (18.5,0);
		\draw[occ,color=colort2] (18.5,0) -- (20,0);
		\draw[occ,color=colort3] (7.5,1) -- (8.5,1);

		\begin{scope}[shift={(0,-2-0.25)}]

		\node [anchor=west]at (-1,1.3) {State 1};
		\draw[rail] (0,0) \railNSend -- (1,0) \railNSend -- (2.5,0) \railNSend -- (3.5,0) \railNSend -- (5,0) \railNSend ;
		\draw[rail] (1.25,0) -- (2.25,1) -- (2.5,1) \railNSend -- (3.5,1) \railNSend -- (3.75,1) -- (4.75,0);
	\begin{scope}[shift={(5,0)}]
		\draw[rail] (0,0) \railNSend -- (1,0) \railNSend -- (2.5,0) \railNSend -- (3.5,0) \railNSend -- (5,0) \railNSend ;
		\draw[rail] (1.25,0) -- (2.25,1) -- (2.5,1) \railNSend -- (3.5,1) \railNSend -- (3.75,1) -- (4.75,0);
	\end{scope}
	\begin{scope}[shift={(10,0)}]
		\draw[rail] (0,0) \railNSend -- (1,0) \railNSend -- (2.5,0) \railNSend -- (3.5,0) \railNSend -- (5,0) \railNSend ;
		\draw[rail] (1.25,0) -- (2.25,1) -- (2.5,1) \railNSend -- (3.5,1) \railNSend -- (3.75,1) -- (4.75,0);
	\end{scope}
	\begin{scope}[shift={(15,0)}]
		\draw[rail] (0,0) \railNSend -- (1,0) \railNSend -- (2.5,0) \railNSend -- (3.5,0) \railNSend -- (5,0) \railNSend ;
		\draw[rail] (1.25,0) -- (2.25,1) -- (2.5,1) \railNSend -- (3.5,1) \railNSend -- (3.75,1) -- (4.75,0);
	\end{scope}
	\draw[rail, ->, >=latex, dashed] (0,0) -- (-1,0); 
		\draw[rail, ->, >=latex, dashed] (20,0) -- (21,0);
		\trainoverarrow{0.1+0.5+1}{0}{1.8}{1.0} 
		\trainoverarrowleft{8.4-5}{1}{0.8}{0.7}
		\trainoverarrowleft{19.4-5}{0}{1.8}{1.0}
		
		\draw[occ,color=colort1] (0,0) -- (1,0);
		\draw[occ,color=colort1] (1,0) -- (2.5,0);
		\draw[occ,color=colort1] (2.5,0) -- (3.5,0);

		\draw[occ,color=colort2] (17.5,0) -- (18.5,0);
		\draw[occ,color=colort2] (18.5,0) -- (20,0);

		\draw[occ,color=colort2] (16,0) -- (17.5,0);
		\draw[occ,color=colort2] (15,0) -- (16,0);
		\draw[occ,color=colort2] (13.5,0) -- (15,0);
		\draw[occ,color=colort2] (12.5,0) -- (13.5,0);	
		
		\draw[occ,color=colort3] (7.5,1) -- (8.5,1);
		\draw[occ,color=colort3] (6,0) -- (6.25,0) -- (7.25,1) -- (7.5,1);
		\draw[occ,color=colort3] (5,0) -- (6,0);
		\draw[occ,color=colort3] (5,0) -- (4.75,0) -- (3.75,1) -- (3.5,1);
		\draw[occ,color=colort3] (2.5,1) -- (3.5,1);

		


		\end{scope}
		\begin{scope}[shift={(0,-4-0.5)}]
			\node [anchor=west]at (-1,1.3) {State 2};
	
			\draw[rail] (0,0) \railNSend -- (1,0) \railNSend -- (2.5,0) \railNSend -- (3.5,0) \railNSend -- (5,0) \railNSend ;
			\draw[rail] (1.25,0) -- (2.25,1) -- (2.5,1) \railNSend -- (3.5,1) \railNSend -- (3.75,1) -- (4.75,0);
		\begin{scope}[shift={(5,0)}]
			\draw[rail] (0,0) \railNSend -- (1,0) \railNSend -- (2.5,0) \railNSend -- (3.5,0) \railNSend -- (5,0) \railNSend ;
			\draw[rail] (1.25,0) -- (2.25,1) -- (2.5,1) \railNSend -- (3.5,1) \railNSend -- (3.75,1) -- (4.75,0);
		\end{scope}
		\begin{scope}[shift={(10,0)}]
			\draw[rail] (0,0) \railNSend -- (1,0) \railNSend -- (2.5,0) \railNSend -- (3.5,0) \railNSend -- (5,0) \railNSend ;
			\draw[rail] (1.25,0) -- (2.25,1) -- (2.5,1) \railNSend -- (3.5,1) \railNSend -- (3.75,1) -- (4.75,0);
		\end{scope}
		\begin{scope}[shift={(15,0)}]
			\draw[rail] (0,0) \railNSend -- (1,0) \railNSend -- (2.5,0) \railNSend -- (3.5,0) \railNSend -- (5,0) \railNSend ;
			\draw[rail] (1.25,0) -- (2.25,1) -- (2.5,1) \railNSend -- (3.5,1) \railNSend -- (3.75,1) -- (4.75,0);
		\end{scope}
		\draw[rail, ->, >=latex, dashed] (0,0) -- (-1,0); 
			\draw[rail, ->, >=latex, dashed] (20,0) -- (21,0);
	
			\trainoverarrow{11.6+5-5}{1}{1.8}{1.0} 
			\trainoverarrowleft{8.4-5}{1}{0.8}{0.7}
			\trainoverarrowleft{19.4-5}{0}{1.8}{1.0}

			\draw[occ,color=colort1] (1,0) -- (2.5,0);
			\draw[occ,color=colort1] (2.5,0) -- (3.5,0);
			\draw[occ,color=colort1] (3.5,0) -- (5.0,0);
			\draw[occ,color=colort1] (5.0,0) -- (6.0,0);
			\draw[occ,color=colort1] (6.0,0) -- (7.5,0);
			\draw[occ,color=colort1] (7.5,0) -- (8.5,0);
			\draw[occ,color=colort1] (8.5,0) -- (10,0);
			\draw[occ,color=colort1] (10,0) -- (11,0);
			\draw[occ,color=colort1] (16-5,0) -- (16.25-5,0) -- (17.25-5,1) -- (17.5-5,1);
			\draw[occ,color=colort1] (17.5-5,1) -- (18.5-5,1);
	
	
			\draw[occ,color=colort2] (13.5,0) -- (15,0);
			\draw[occ,color=colort2] (12.5,0) -- (13.5,0);		
			
			\draw[occ,color=colort3] (7.5-5,1) -- (8.5-5,1);

			\end{scope}
		\begin{scope}[shift={(0,-6-0.75)}]
		\node [anchor=west]at (-1,1.3) {State 3};

		\draw[rail] (0,0) \railNSend -- (1,0) \railNSend -- (2.5,0) \railNSend -- (3.5,0) \railNSend -- (5,0) \railNSend ;
		\draw[rail] (1.25,0) -- (2.25,1) -- (2.5,1) \railNSend -- (3.5,1) \railNSend -- (3.75,1) -- (4.75,0);
	\begin{scope}[shift={(5,0)}]
		\draw[rail] (0,0) \railNSend -- (1,0) \railNSend -- (2.5,0) \railNSend -- (3.5,0) \railNSend -- (5,0) \railNSend ;
		\draw[rail] (1.25,0) -- (2.25,1) -- (2.5,1) \railNSend -- (3.5,1) \railNSend -- (3.75,1) -- (4.75,0);
	\end{scope}
	\begin{scope}[shift={(10,0)}]
		\draw[rail] (0,0) \railNSend -- (1,0) \railNSend -- (2.5,0) \railNSend -- (3.5,0) \railNSend -- (5,0) \railNSend ;
		\draw[rail] (1.25,0) -- (2.25,1) -- (2.5,1) \railNSend -- (3.5,1) \railNSend -- (3.75,1) -- (4.75,0);
	\end{scope}
	\begin{scope}[shift={(15,0)}]
		\draw[rail] (0,0) \railNSend -- (1,0) \railNSend -- (2.5,0) \railNSend -- (3.5,0) \railNSend -- (5,0) \railNSend ;
		\draw[rail] (1.25,0) -- (2.25,1) -- (2.5,1) \railNSend -- (3.5,1) \railNSend -- (3.75,1) -- (4.75,0);
	\end{scope}
	\draw[rail, ->, >=latex, dashed] (0,0) -- (-1,0); 
		\draw[rail, ->, >=latex, dashed] (20,0) -- (21,0);

		\trainoverarrow{11.6+5-5}{1}{1.8}{1.0} 
		\trainoverarrowleft{0.9}{0}{0.8}{0.7}
		\trainoverarrowleft{19.4-5}{0}{1.8}{1.0}
		
		\draw[occ,color=colort1] (16-5,0) -- (16.25-5,0) -- (17.25-5,1) -- (17.5-5,1);
		\draw[occ,color=colort1] (17.5-5,1) -- (18.5-5,1);


		\draw[occ,color=colort2] (13.5,0) -- (15,0);
		\draw[occ,color=colort2] (12.5,0) -- (13.5,0);		
		
		\draw[occ,color=colort3] (7.5-5,1) -- (8.5-5,1);
		\draw[occ,color=colort3] (0,0) -- (1,0);
		\draw[occ,color=colort3] (6-5,0) -- (6.25-5,0) -- (7.25-5,1) -- (7.5-5,1);

		\end{scope}
			 	
		\begin{scope}[shift={(0,-8-1)}]
					\node [anchor=west]at (-1,1.3) {State 4 -- {\color{red!75!black}\textsc{Deadlocked}}};
				 \end{scope}

\end{scope}
	\end{tikzpicture} 
		}
	\end{center}
\caption{
	The railway infrastructure and trains shown in State 0 is bound for a deadlock, because
	the left-most and the right-most trains are too long to pass each other anywhere.
	States 1-3 are an example from the set of the longest possible plans 
	for this system when applying the partial order reduction described in this paper.
	Note, for example, that the right-most train moves already in State 1 to the position where it will
	eventually block the left-most train. 
	Using a pre-computed upper bound on the number of steps would instead require 26 transitions,
	making the resulting SAT instance harder to solve.
}
\label{fig:deadlockrail:introexampleimproved}

\end{figure}

%% file: routefig.tex
\begin{figure}[b]
  \centering

    \begin{tikzpicture}
        \pic at (0,0) {station};
        \node at (0.0,-0.6) {A};
        \node at (5.0,-0.6) {C};

        \draw[yshift=0.5em, line width=0.4em,-{latex},color=green!50!black,opacity=0.8] (0,0) -- (5,0);
        \draw[yshift=0.5em, line width=0.4em, dashed, color=green!50!black,opacity=0.8] (1.5,0) -- (2.5,1) -- (3,1);


    \end{tikzpicture}

%
%
%
%

\caption{Elementary route AC from signal A to the adjacent signal C. 
The thick line indicates parts of the track on the train's path which are 
reserved for this movement, and the dashed lines indicate 
    parts of the track outside the path which are also exclusively allocated.}
\label{fig:scheduling:elementaryroute}
\end{figure}

%% file: plannerinputstructs.tex
\begin{figure}
    \begin{center}
	    \begin{tikzpicture}[xscale=1.3,yscale=1.3]

                    \draw[fill=black] (1.5,0) -- (1.75,0) -- (1.75,0.25);
                    \draw[fill=black] (5.5,0) -- (5.75,0) -- (5.75,-0.25);
                    \draw[fill=black] (7.5,0) -- (7.25,0) -- (7.25,0.25);

                    \draw[rail] (-0.5,0) -- (0,0) \railNSend -- (1,0) \railNSend -- (1.5,0);
                    \draw[rail] (1.5,0) -- (2.5,1) -- (3.0,1) \railNSend -- (5.0,1) \railNSend -- (6.0,1) \railNSend -- (6.5,1) -- (7.5,0);
                    \draw[rail] (1.5,0) -- (3.0,0) \railNSend -- (5.0,0) \railNSend -- (5.5,0);
                    \draw[rail] (5.5,0) -- (6.5,-1) -- (7,-1) \railNSend -- (9,-1);
                    \draw[draw=none,fill=black] (9.0,-0.8) rectangle (9.1,-1.2);

                    \draw[rail] (5.5,0) -- (6.0,0) \railNSend -- (8,0) \railNSend -- (9,0) \railNSend -- (9.5,0);

                    \draw[->,>=latex,ultra thick] (-0.5,0) --(-0.9,0);
                    \draw[->,>=latex,ultra thick] (9.5,0) --(9.9,0);

                    \signalsmall{0}{-0.3}
                    \signalsmall{5}{-0.3}
                    \signalsmall{5}{1.3}
                    \begin{scope}[rotate=180]
                    \signalsmall{-3}{-1.3}
                    \signalsmall{-3}{0.3}
                    \signalsmall{-9}{-0.3}
                    \signalsmall{-7}{1.0-0.3}
                    \end{scope}

\draw [yshift=0.5em, line width=0.4em, -{latex},color=green!50!black,opacity=0.8] (-0.0,0) -- node[near end,above,color=green!30!black,opacity=1,xshift=1em]{$\left\{ r_1, r_2, r_3 \right\} \in \text{ElemRoutes}$} (5.0,0) ;
\draw [yshift=-2.0em, line width=0.4em, -{latex},color=blue!50!black,opacity=0.8] (-0.0,0) -- node[midway,below,xshift=-2.5em,yshift=-0.5em,color=blue!30!black,opacity=1]{$r_1 \in \text{PRoutes}$} (1.0,0) ;
\draw [yshift=-2.0em, line width=0.4em, -{latex},color=blue!50!black,opacity=0.8] (1.0,0) -- node[midway,below,yshift=-0.5em,color=blue!30!black,opacity=1]{$r_2 \in \text{PRoutes}$} (3.0,0) ;
\draw [yshift=-2.0em, line width=0.4em, -{latex},color=blue!50!black,opacity=0.8] (3.0,0) -- node[midway,below,xshift=0.75em,yshift=-0.5em,color=blue!30!black,opacity=1]{$r_3 \in \text{PRoutes}$} (5.0,0) ;

 \draw[thick, decoration={brace,mirror,raise=8pt},decorate,yshift=-1em]
         (0,-1) -- node[below=1em,xshift=-0.75em] {$\text{routeLength}(r_1) = 50.0$} (0.9,-1);
 \draw[thick, decoration={brace,mirror,raise=8pt},decorate,yshift=-1em]
         (1.1,-1) -- node[below=1em+1.1em] {$\text{routeLength}(r_2) = 50.0$} (2.9,-1);
 \draw[thick, decoration={brace,mirror,raise=8pt},decorate,yshift=-1em]
         (3.1,-1) -- node[below=1em, xshift=0.75em] {$\text{routeLength}(r_3) = 250.0$} (5,-1);

         \draw[line width=5pt,white](1,1.5) -- (1,0.3);
         \draw[thick, red!50!black, ->,>=latex] (1,1.5) -- node[anchor=south,at start,xshift=2em]
         {$\text{exit}(r_1) = \text{entry}(r_2) = d_2 \in \text{Delims}$} (1,0.3);

         \draw[line width=5pt,white](0,0.75) -- (0,0.3);
         \draw[thick, red!50!black, ->,>=latex] (0,0.75) -- node[anchor=south,at start,fill=white,xshift=-2.5em]
         {$\text{entry}(r_1) = d_1 \in \text{Delims}$} (0,0.3);

         \node[anchor=west,red!50!black] at (6.35,-1.6) {$\text{exit}(r_3)=d_4 \in \text{Delims}$};
         \draw[thick, red!50!black,->,>=latex] (6.35,-1.6) -- (5.25,-0.5);

    \end{tikzpicture}
    \end{center}
    \caption{The abstracted infrastructure input used to represent
     the infrastructure in the planning SAT problem. The elementary route 
    AC (see \figref{fig:scheduling:elementaryroute}) 
    consists of three partial routes, $r_1, r_2, r_3$, connected 
    through their shared delimiters ($\text{exit}(r_1) = \text{entry}(r_2)$, etc.).
     The exit signal  of the elementary route is the delimiter $d_4$ 
     (corresponding to signal C in \figref{fig:scheduling:elementaryroute})
    }
    \label{fig:scheduling:plannerinputstructs}
\end{figure}

%% file: twotraintable.tex
{
\newcommand{\timeout}{{\scriptsize $>$60.00}}
\setlength{\tabcolsep}{1.0em} 
\begin{table}
\centering
\begin{tabular}{rr|rr|rr|rr}
\hline
\multicolumn{2}{c}{\small Instance}                              & \multicolumn{2}{c}{\small Algorithm 1}                        & \multicolumn{2}{c}{\small Algorithm 2}                          & \multicolumn{2}{c}{\small Algorithm 3}                      \\ 
\multicolumn{1}{c}{\small Stations} & \multicolumn{1}{c}{\small Routes} & \multicolumn{1}{c}{\small Steps} & \multicolumn{1}{c}{\small Time (s)}   & \multicolumn{1}{c}{\small Steps} & \multicolumn{1}{c}{\small Time (s)}     & \multicolumn{1}{c}{\small Steps} & \multicolumn{1}{c}{\small Time (s)} \\ \hline
2                            & 16                         & 10 (UB)                   & 0.00                       & 7                        & 0.00                           &    3                      &   0.00                       \\
4                            & 32                         & 22 (UB)                   & 0.15                       & 13                       & 0.19                         &      3                    &     0.00                     \\
6                            & 48                         & 34 (UB)                   & 2.80                       & 19                       & 1.70                         &      3                    &     0.00                     \\ 
8                            & 64                         & 46 (UB)                   & 18.90                      & 25                       & 6.70                           &    3                      &   0.00                       \\
10                           & 80                         & -                         & \timeout                   & 31                       & 35.90                           &    3                      &   0.00                       \\
20                           & 160                        & -                         & \timeout                   & -                         & \timeout                         &   3                       &  0.01                        \\
50                           & 400                        & -                         & \timeout                   & -                         & \timeout                         &   3                       &  0.04                        \\
100                          & 800                        & -                         & \timeout                   & -                         & \timeout                         &   3                       &  0.28                       \\ \hline
\end{tabular}
\caption{Scaling test with two trains headed in opposite directions on a sequence of two-track stations. Each station track is too short to contain any train, so the situation is bound for deadlock. 
         Each row describes a problem instance with a number of two-track stations and corresponding number of routes in the problem. 
	 For each of the algorithms 1, 2, 3, two columns show  the number of planning steps added to the problem before a conclusion is reached, and the running time of the algorithm. A timeout of 60 seconds was used.
	}
\label{tab:deadlockrail:twotrainscaling}
\end{table}
}

%% file: perftable.tex
{
\newcommand{\timeout}{{\scriptsize $>$60.00}}
\newcommand{\dead}{{\small \textsc{Dead}}}
\newcommand{\live}{{\small \textsc{Live}}}
\setlength{\tabcolsep}{0.5em} 
\begin{table}
\centering
	{
\begin{tabular}{llrr|rr|rr|rr|rr}
\hline
	\multicolumn{4}{c}{\small Instance}                                                            & \multicolumn{2}{c}{\small Ticks MIP alg.}                        & \multicolumn{2}{c}{\small Algorithm 1}                        & \multicolumn{2}{c}{\small Algorithm 2}                          & \multicolumn{2}{c}{\small Algorithm 3}                      \\ 
	\multicolumn{4}{c}{}                                                            & \multicolumn{2}{c}{\small (reported in \cite{sassoticks})}                        & \multicolumn{2}{c}{}                        & \multicolumn{2}{c}{}                          & \multicolumn{2}{c}{}                      \\ 
	\multicolumn{1}{c}{\small \#} &\multicolumn{1}{c}{\small Result} & \multicolumn{1}{c}{\small $n_r$}& \multicolumn{1}{c}{\small $n_t$} & \multicolumn{1}{c}{\small Steps} & \multicolumn{1}{c}{\small Time (s)} & \multicolumn{1}{c}{\small Steps} & \multicolumn{1}{c}{\small Time (s)}   & \multicolumn{1}{c}{\small Steps} & \multicolumn{1}{c}{\small Time (s)}     & \multicolumn{1}{c}{\small Steps} & \multicolumn{1}{c}{\small Time (s)} \\ \hline
	01 & \live &   14  & 3  & 8  &1.08       & 5  & 0.00 & 5  & 0.00     & 5 & 0.00 \\
	02 & \dead &   14  & 3  & 8  &0.98       & 10 & 0.00 & 7  & 0.00     & 5 & 0.00 \\
	03 & \live &   14  & 3  & 8  &0.93       & 5  & 0.00 & 5  & 0.00     & 5 & 0.00 \\
	04 & \live &   30  & 2  & 15 &1.20       & 4  & 0.00 & 4  & 0.00     & 4 & 0.00 \\
	05 & \live &   30  & 3  & 20 &1.31       & 5  & 0.00 & 5  & 0.00     & 5 & 0.00 \\
	06 & \dead &   30  & 3  & 20 &2.78       & 19 & 0.03 & 9  & 0.04     & 5 & 0.00 \\
	07 & \dead &   38  & 5  & 34 &37.31      & 34 & 0.17 & 7  & 0.00     & 5 & 0.00 \\
	08 & \live &   46  & 5  & 33 &1.78       & 5  & 0.00 & 5  & 0.00     & 5 & 0.00 \\
	09 & \dead &   38  & 6  & 37 &\timeout   & 37 & 0.26 & 14 & 0.25     & 7 & 0.01 \\
	10 & \dead &   38  & 7  & 42 &4.23       & 42 & 0.02 & 2  & 0.00     & 2 & 0.00 \\
	11 & \dead &   62  & 2  & 27 &17.60      & 26 & 3.30 & 15 & 3.30     & 3 & 0.00 \\
	12 & \dead &   62  & 4  & 39 &\timeout   & 40 & 1.70 & 20 & 9.60     & 8 & 0.19 \\
	13 & \dead &   62  & 4  & 39 &\timeout   & 40 & 1.30 & 20 & 11.00    & 8 & 0.12 \\
	14 & \live &   62  & 4  & 39 &3.27       & 6  & 0.00 & 6  & 0.01     & 6 & 0.02 \\
	15 & \dead &   46  & 4  & 42 &\timeout   & 42 & 0.22 & 15 & 1.30     & 6 & 0.01 \\
	16 & \live &   62  & 5  & 50 &5.33       & 5  & 0.00 & 5  & 0.00     & 5 & 0.01 \\
	17 & \live &   62  & 4  & 50 &43.11      & 6  & 0.00 & 6  & 0.01     & 6 & 0.02 \\
	18 & \dead &   62  & 4  & 50 &\timeout   & 49 & 2.10 & 15 & 13.10    & 6 & 0.05 \\
	19 & \dead &   62  & 5  & 51 &\timeout   & 50 & 0.83 & 16 & 36.00    & 6 & 0.03 \\
	20 & \dead &   70  & 5  & 57 &\timeout   & 56 & 1.20 & -  & \timeout & 6 & 0.03 \\
\hline
\end{tabular}
	}
\caption{
	Performance evaluation on the instances from \cite{sassoticks}.
	The $n_r$ and $n_t$ columns show the number of routes and number of trains, respectively, 
	indicating the size of the problem instance.
	The ``Ticks MIP alg.'' column refers to the algorithm in \cite{sassoticks} and 
	relays the worst-case results reported in that paper.
	For algorithms 1, 2, and 3, the ``Steps'' columns contains to the number
	of transitions that have been added to the SAT formula when the algorithm terminates,
	and ``Time'' column reports the running time in seconds.
	}
\label{tab:deadlockrail:perftable}
\end{table}
}